\documentclass{mn2e}
\usepackage{psfig}
\def\simgt{\hbox{\rlap{\raise 0.425ex\hbox{$>$}}\lower 0.65ex\hbox{$\sim$}}}
\def\simlt{\hbox{\rlap{\raise 0.425ex\hbox{$<$}}\lower 0.65ex\hbox{$\sim$}}}

\def\bj {b_{\rm J}}

\def\kmsmpc {km$\,$s$^{-1}$Mpc$^{-1}$}
\def\om {\Omega_{\rm m}}
\def\ol {\Omega_{\Lambda}}
\def \ho {H_0}
\def \oiii {[O{\small~III}]}

\def \nq {N_{\rm Q}}

\def \lp {\log\Phi}
\def \dlp {\Delta\log\Phi}

\def \mg {M_{\rm g}}
\def \mi {M_{\rm i}}

\def \kem {K_{\rm em}}

\def \mbj {M_{\rm b_{\rm J}}}
\def \mg {M_{\rm g}}
\def \kev {keV}
\def \zc {z_{\rm c}}
\def \zcz {z_{\rm c,0}}
 
\def \aj {AJ}
\def \mnras {MNRAS}
\def \apj {ApJ}
\def \apjs {ApJS}

\def \aap {A\&A}

\title[The 2SLAQ QSO luminosity function]{The 2dF-SDSS LRG and QSO
  Survey: The QSO luminosity function at $0.4<z<2.6$.}

\author[S.~M. Croom et al.]
{Scott M. Croom$^{1}$\thanks{scroom@physics.usyd.edu.au},
Gordon T. Richards$^{2}$,
Tom Shanks$^{3}$,
Brian J. Boyle$^{4}$,
\newauthor
Michael A. Strauss$^{5}$,
Adam D. Myers$^{6}$,
Robert C. Nichol${^7}$,
Kevin A. Pimbblet$^{8}$,
\newauthor
Nicholas P. Ross$^{3,9}$,
Donald P. Schneider$^{9}$,
Robert G. Sharp$^{10}$,
David A. Wake$^{3}$  \\ ${^1}$ Sydney Institute for Astronomy, School of
Physics, University of Sydney, NSW 2006, Australia\\ ${^2}$ Drexel
University, Department of Physics, Philadelphia, PA 19104, USA\\
$^{3}$ Department of Physics, University of Durham, South Road, Durham
DH1 3LE\\ $^{4}$ Australia Telescope National Facility, PO Box 76,
Epping NSW 1710, Australia\\ $^{5}$ Princeton University Observatory,
Peyton Hall, Princeton, NJ 08544, USA\\ $^{6}$ Department of
Astronomy, University of Illinois at Urbana-Champaign, Urbana, IL
61801\\ $^{7}$ Institute of Cosmology and Gravitation, Mercantile
House, Hampshire Terrace, University of Portsmouth, Portsmouth, PO1
2EG\\ $^{8}$ Department of Physics, University of Queensland,
Brisbane, QLD 4072, Australia\\ $^{9}$ Department of Astronomy and
Astrophysics, 525 Davey Laboratory, Pennsylvania State University,
University Park, PA 16802.\\
$^{10}$ Anglo-Australian Observatory, PO Box 296, Epping, NSW 1710,
Australia }

\begin{document}

\maketitle

\newcommand{\fmmm}[1]{\mbox{$#1$}}
\newcommand{\scnd}{\mbox{\fmmm{''}\hskip-0.3em .}}
\newcommand{\scnp}{\mbox{\fmmm{''}}}

\begin{abstract}

We present the QSO luminosity function of the completed 2dF-SDSS LRG
and QSO (2SLAQ) survey, based on QSOs photometrically selected from
Sloan Digital Sky Survey imaging data and then observed
spectroscopically using the 2dF instrument on the  Anglo-Australian
Telescope.  We analyse 10637 QSOs in the redshift range $0.4<z<2.6$ to
a $g$-band flux limit of 21.85 (extinction corrected) and an absolute
continuum magnitude of $\mg(z=2)<-21.5$.  This sample covers an area
of 191.9~deg$^2$.

The binned QSO luminosity function agrees with that of the brighter
SDSS main QSO sample, but extends $\sim2.5$ mags fainter, clearly
showing the flattening of the luminosity function towards faint
absolute magnitudes.  2SLAQ finds an excess of QSOs compared to the
2dF QSO Redshift Survey at $g>20.0$, as found previously by Richards
et al.\ (2005).  The luminosity function is consistent with other
previous, much smaller, samples produced to the depth of 2SLAQ.

By combining the 2SLAQ and SDSS QSO samples we produce a QSO
luminosity function with an unprecedented combination of precision and
dynamic range.  With this we are able to accurately constrain both the
bright and faint ends of the QSO LF.  While the overall trends seen in
the evolution of the QSO LF appear similar to pure luminosity
evolution, the data show very significant departures from such a
model.  Most notably we see clear evidence that the number density of
faint QSOs peaks at lower redshift than bright QSOs: QSOs with
$\mg>-23$ have space densities which peak at $z<1$, while QSOs at
$\mg<-26$ peak at $z>2$.  By fitting simple luminosity function models
in narrow $\mg$ intervals we find that this {\it downsizing} is
significant at the 99.98 per cent level.

We show that luminosity function models which follow the pure
luminosity evolution form [i.e. $\mg^*\equiv\mg^*(z)$], but with a
redshift--dependent bright end slope and an additional density
evolution term, $\Phi^*\equiv\Phi^*(z)$, provide a much improved fit
to the data.  The bright end slope, $\alpha$, steepens from
$\alpha\simeq-3.0$ at $z\simeq0.5$ to $\alpha=-3.5$ at $z\simeq2.5$.
This steepening is significant at the 99.9 per cent level.  We find a
decline in $\Phi^*$ from $z\simeq0.5$ to $z\simeq2.5$ which is
significant at the 94 per cent level.
\end{abstract}


\section{Introduction}

Accurate measurement of the luminosity function is of prime importance
in the study of active galactic nuclei (AGN).  One of the key goals of
studying AGN is to 
characterize and understand their strong evolution (e.g. Schmidt 1972;
Braccesi et al. 1980; Schmidt \& Green 1983; Boyle, Shanks
\& Peterson 1988; Hewett,  Foltz \& Chaffee 1993; Boyle et al. 2000;
Croom et al. 2004, henceforth C04; Richards et al. 2006, henceforth R06).
Early measurements of the QSO luminosity function (LF) demonstrated
strong evolution in the population, with luminous QSOs being much more
common at high redshift ($z\sim2$).  However, because the shape (at bright
magnitudes) is a featureless power law, the type of evolution could not
be quantified,  i.e., there was no observable difference between
density evolution (a change in the number of objects) and luminosity
evolution (a change in the luminosities of objects).  Surveys
which probed fainter than the observed break in the LF (e.g. Boyle et
al 1990) started to allow some understanding of the physical process
behind QSO evolution.  These studies showed that QSO evolution approximately
followed pure luminosity evolution (PLE), with the same characteristic
LF shape evolving to higher luminosities at higher redshift.  A naive
interpretation of such evolution would imply that QSOs are
cosmologically long--lived and fade from $z\sim2$ to $z\sim0$.  Given
their low space density relative to normal galaxies, this
also implies that QSOs are intrinsically rare.  However, more
accurate measurements of the bright end of the QSO LF showed an
evolving slope (Hewett et al. 1993; Goldschmidt \& Miller 1998; R06),
suggesting that QSO evolution did not perfectly follow the PLE model.
At redshifts $\simgt2.5$ there is an observed decline in the space
density of bright QSOs (Osmer 1982; Warren, Hewett \& Osmer 1994;
Schmidt, Schneider \& Gunn 1995; Fan et
al. 2001).  Thus $z\simeq2-3$ is often known as the {\it quasar
  epoch}, where QSOs (or quasars) were most active. 

The realization that most massive galaxies contain super-massive
black holes (SMBHs) (e.g. Kormendy \& Richstone 1995) meant that QSOs were
likely to be intrinsically common.  This result is much more suggestive of a
model where QSOs are a short-lived process that occurs in most massive
galaxies.  The observed evolution is then due to global effects such
as a decline in the triggering rate or fuelling of AGN, which
modulates the distribution of many successive generations.

Early measurements of the X-ray AGN LF (e.g. Boyle et al. 1993) showed
evolution which also approximately followed PLE.  However, the most
recent X-ray surveys, in particular using {\it Chandra} (e.g. Giacconi
et al. 2002; Alexander et al. 2003) and {\it XMM-Newton} 
(e.g. Hasinger et al. 2001; Worsley et al. 2004), reach to much
fainter flux levels.  These surveys have demonstrated that pure
luminosity evolution fails to match the redshift dependent evolution
of the AGN LF at $L<L^*$ (where $L^*$ is the characteristic luminosity
at which the QSO LF flattens).  They show that the activity in faint AGN
peaks at a lower redshift than that of more luminous AGN
(e.g. Ueda et al. 2003; Hasinger et al 2005);  this process has been described as AGN
downsizing  (e.g. Barger et al. 2005).  The galaxy population is also
seen to undergo a similar downsizing (Cowie et al. 1996), where the
most massive galaxies formed the bulk of their stars earlier than lower mass
galaxies (e.g. Juneau et al. 2005; Zheng et al. 2007; Noeske et
al. 2007).

Given that all massive galaxies contain a SMBH,
and that there are tight correlations between black hole mass and host
properties (e.g. Tremaine et al. 2002), the growth of black holes and
galaxies must be intimately connected.  In particular, it has been
proposed that galaxy mergers trigger major episodes of star formation
(e.g. in ultra-luminous infrared galaxies) and lead to QSO activity
(Sanders et al. 1988).  Such a mechanism can plausibly form spheroidal
galaxies and QSOs and this idea has been further developed by recent numerical
simulations (e.g. Hopkins et al. 2005a).  In such a picture, accretion
onto a SMBH is triggered (at least for moderate to high luminosity AGN)
by the merger of gas-rich galaxies.  While the timescale for the merger may be
$\sim1$Gyr, for the majority of this time the accretion is obscured from
view by dust.  Only when the AGN finally expels the surrounding gas and
dust does it shine as an optical QSO for a brief period ($\sim100$Myr),
before exhausting its fuel supply (e.g. Di Matteo et al. 2005).  This
feedback of energy from the AGN into the host also heats and expels the gas in the
galaxy, which suppresses star formation leading to ``red and dead''
ellipticals or bulges.  One of the key predictions of the Hopkins et
al. model is that the faint end of the QSO luminosity function should
largely be comprised of high mass SMBHs at low accretion rates
(i.e. well below their peak luminosity) rather than lower mass SMBHs accreting near the
Eddington rate \cite{hop05b}.
Thus, while the bright end of the QSO LF tells us about the intrinsic
properties of the QSO population during the time when black holes
where increasing in mass most rapidly (e.g. triggering rate, active black
hole mass function etc.), the faint end of the LF tells us about the
length of time QSOs spend at relatively low accretion rates.

The 2-degree Field (2dF) QSO Redshift Survey (2QZ; Croom et al. 2001; 2004)
covered an area of $720$ sq deg, and reached $\sim1$ mag
fainter than the break in the QSO LF at $z<2$.  However, as the
observed break in the LF is a relatively gradual flattening towards
faint magnitudes, the constraints from the 2QZ on the actual slope of
the faint end are uncertain, as demonstrated by the difference
between the results from the first release (Boyle et al. 2000) and the
final release (Croom et al. 2004) of the 2QZ.  Samples that reach substantially deeper are required to
properly constrain the shape of the faint QSO LF.  Such deeper
spectroscopic surveys have only covered small areas to date.  Wolf et al. (2003) used
the medium-band photometric data from the COMBO-17 survey to construct
a QSO LF to $R<24$ that contains $\sim200$ QSOs over an area of 1
deg$^2$.  They provided a measurement of the QSO LF over the redshift 
range $1.2<z<4.8$, but were unable to differentiate between density and
luminosity evolution.  Jiang et al. (2006) used the deep Sloan Digital
Sky Survey (SDSS; York et al. 2000) data from the Fall Equatorial
Stripe (``Stripe 82'') to construct a sample of 400 QSOs over $4$ deg$^2$ to
$g<22.5$.  They found good agreement with the early 2dF-SDSS LRG And
QSO (2SLAQ) survey results of
Richards et al. (2005; henceforth R05), but were not able to see any clear evidence of
downsizing in the AGN population.  Using data from the VLT VIMOS Deep Survey
(VVDS), Bongiorno et al. (2007) constructed a QSO LF to a limit of $I_{\rm
  AB}<24$ with 130 QSOs in an area of 1.7 deg$^2$.  When combined with
the SDSS data from R06, these authors found that
their data are better fit by a luminosity--dependent density evolution
model (LDDE), which also matches X-ray samples, suggestive of downsizing.

The 2SLAQ survey (Croom et al. 2009; C09) was specifically designed to
probe the faint end of the QSO LF, reaching approximately 1 magnitude deeper than the
2QZ.  A key requirement of the 2SLAQ survey was that it should cover sufficient
area to allow accurate 
measurement of QSO clustering, as well as to reduce the random errors in
the measurement of the LF, spectral properties etc.  Richards et
al. (2005) presented the QSO luminosity function from an early 2SLAQ
data set using $\simeq5600$ QSOs; this result shows an excess over the
2QZ survey at $g\simeq21$,
but still clearly demonstrates a break in the LF.  In this paper we
present the QSO luminosity function for the final 2SLAQ sample.  In
addition to containing approximately twice as many QSOs, this analysis also makes
use of the improved completeness estimates presented by C09, which
includes the impact of QSO host galaxies. In
Section \ref{sec:data} we briefly describe the 2SLAQ survey.
In Section \ref{sec:kcorr} we discuss in detail the issue of
K-corrections, followed by a calculation of QSO number counts in
Section \ref{sec:nm}.  Binned estimates of the QSO LF
are presented in Section \ref{sec:binnedlf}, including a combined
2SLAQ+SDSS LF, and we describe model fits
to the data in Section \ref{sec:mlfits}.  Section \ref{sec:dis}
discusses our results in the context of recent models for galaxy and
QSO evolution.  We summarize our conclusions
in Section \ref{sec:sum}.  Throughout this paper we assume a
cosmological model with $\ho=70$\kmsmpc, $\om=0.3$ and $\ol=0.7$.  All
photometric measurements quoted in this paper have been corrected for
Galactic extinction using the maps of Schlegel, Finkbeiner \& Davis (1998).

\section{The 2SLAQ Survey}\label{sec:data}

The 2SLAQ survey combines $ugriz$ \cite{fuk96} photometry from SDSS
DR1 (Gunn et al. 1998, 2006; Stoughton et~al. 2002; Abazajian
et~al. 2003) and deep spectroscopy using the 2dF
spectrograph on the Anglo-Australian Telescope (Lewis et al. 2002).
The survey is described in detail by C09.  In this section we
summarize the key properties of the sample.  QSO candidates are
selected with $18.0<g<21.85$ [SDSS point spread function (PSF)
photometry, extinction corrected], using a
multi-colour method which primarily selects UV-excess objects.  This
limits the redshift range to $z\simlt3$, with the 
completeness falling below 50 per cent at $z>2.6$.  QSO candidates
fainter than $g=20.5$ had higher priority when configuring a field for
observation with 2dF, as the main focus of the
survey was on the faint end of the QSO luminosity function.
The 2SLAQ survey covers an area of 191.9 deg$^2$ in two regions along
the celestial equator (declination $=-1.259$ to $+0.840^{\circ}$) in
the North and South galactic caps (henceforth named the NGP and SGP
regions).  This area corresponds to a cosmological volume of 4.0~Gpc$^3$
over the redshift range $0.4<z<2.6$ (in our assumed cosmology).

The observations contain new spectra of 16326 objects, of
which  8764 are QSOs.  A total of 7623 of these are newly discovered,
with the remainder previously identified by the 2QZ (Croom et
al. 2004) and SDSS \cite{sdssqso4} surveys.  The full QSO sample
contains 12702 QSOs and is presented in C09.

C09 discuss the completeness of the 2SLAQ QSO sample in detail.  In
particular, because the survey is somewhat fainter than previous large
QSO surveys, it is important to take into account the effect of the
host galaxy on completeness.  C09 showed that QSOs at $z<1$ near the
faint limit of the survey are significantly redder because of the
contribution of the host galaxy component, and that this reduces the
completeness of the colour selection.  This has also been demonstrated
in the small but complete sample of QSOs from the VVDS (Gavignaud et
al. 2006).

\section{The K-correction}\label{sec:kcorr}

An accurate
K-correction is required to properly account for the redshifting of
the observed pass-bands when calculating absolute magnitudes or
fluxes.  In QSO spectra, broad emission lines can also contribute a
significant fraction of the flux (typically 0.2--0.5 mags) in a
photometric band.  As faint targets in the 2SLAQ
sample can be affected by their host galaxy, we also need to subtract
the host flux to obtain the nuclear component.  This is in contrast to
most previous samples (e.g. 2QZ, SDSS), where an absolute magnitude
limit was applied to the sample, and objects brighter than that limit
were assumed not to be significantly affected by flux from their
hosts.  

\subsection{Correcting for host galaxy flux}

The detailed completeness simulations described by C09 enable
correction of the 2SLAQ QSO magnitudes for their mean host galaxy
contributions.  The photometric data used to select QSOs are the SDSS
PSF magnitudes, so this will already limit the host galaxy flux to
some extent (e.g. Schneider et al. 2003).  However, C09 demonstrated
that at low luminosity the host galaxies of the 2SLAQ
sample can still significantly alter their observed colours.  We use the
modelled host properties presented by C09 to correct the 2SLAQ QSOs
for the contribution of their host galaxies.  For each simulated
source C09 calculate the total, nuclear--only and host--only magnitudes.  We
can then derive the mean correction from total to nuclear magnitude in
$g$ and $z$ intervals (note that here, and throughout this paper, $z$
denotes redshift and not a magnitude in the SDSS $z$-band).  This is
shown in Fig. 16a and Table 12 of C09.  In the 
$g$-band, in which the 2SLAQ sample was selected, the host
contribution is less than 20 per cent at $z>0.4$, even 
for the faintest sources.  Thus, corrections for the host galaxy have
only a limited effect on the measured QSO LF.

\subsection{The QSO K-correction}\label{sec:qsokcorr}

\begin{figure}
\centering
\centerline{\psfig{file=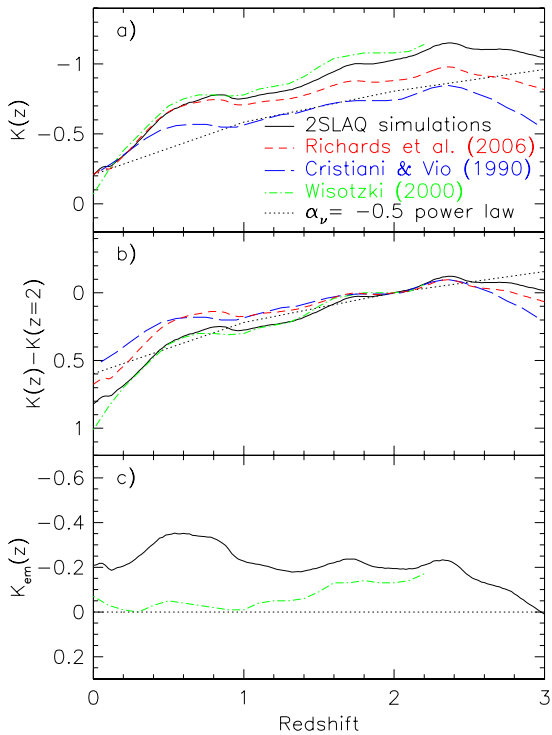,width=8cm}}
\caption{a) The $g$-band QSO K-corrections normalized to the continuum at
  $z=0$ from the 
  2SLAQ simulations of C09 (solid black line), R06 (short--dashed red
  line), Cristiani \& Vio 
  (2000) (long--dashed blue line), Wisotzki (2000) (dot--dashed green line) and a power law
  with $\alpha_\nu=-0.5$ (dotted black line).  The Cristiani
  \& Vio K-correction has been transformed from their $B$- and
  $V$-band measurements to the $g$-band.  The 2SLAQ
  K-correction comprises an emission line component and a power law
  with $\alpha_\nu=-0.3$.  The R06, Cristiani \& Vio and power law 
  K-corrections have been shifted by $-0.209$ to correct for the $z=0$
  emission line contribution.  This has not been done to the Wisotski
  K-correction which already includes an emission line term.  The
  emission line contribution causes these K-corrections not to pass
  through zero at $z=0$, even though they are normalized at $z=0$.  b) The
  same K-corrections, but normalized at $z=2$.  c) The emission line
  K-corrections from our work (black solid line) and Wisotzki (2000) (green
  dot--dashed line).} 
\label{fig:kcorr}
\end{figure}

\begin{figure}
\centering
\centerline{\psfig{file=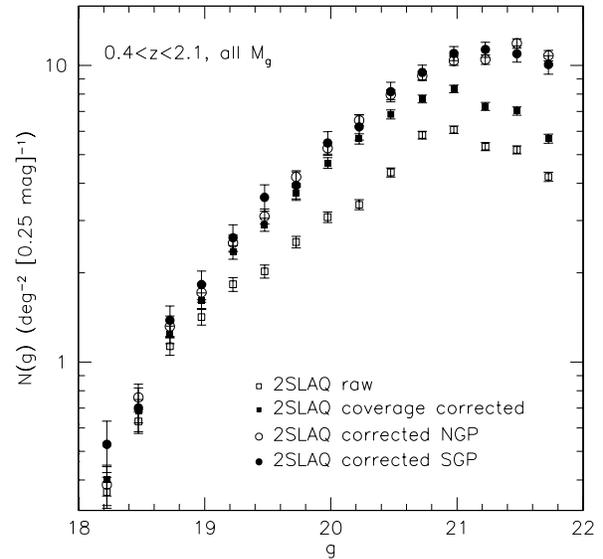,width=8cm}}
\caption{ The $g$-band QSO number counts from the 2SLAQ survey at
  $0.4<z<2.1$ (with no absolute magnitude limits).  In this plot we
  compare the raw counts (open squares) and counts corrected for coverage
  completeness only (filled squares) to the counts after applying all
  corrections (coverage, photometric, spectroscopic etc. see C09) in
  the NGP and SGP regions separately (open and filled circles).}  
\label{fig:nmcomp}
\end{figure}

\begin{figure}
\centering
\centerline{\psfig{file=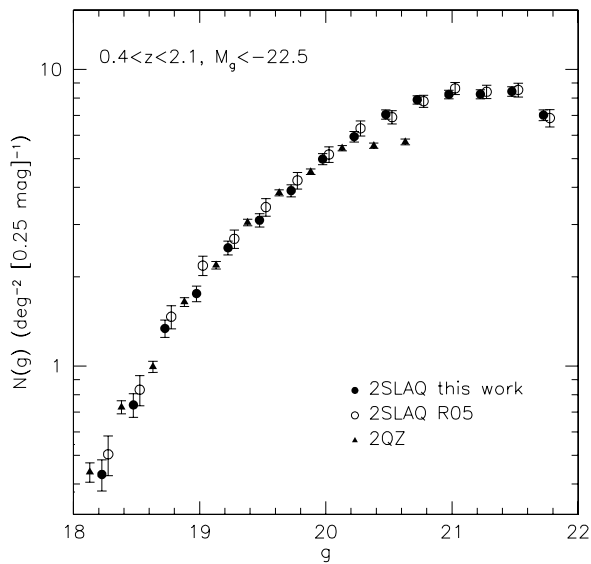,width=8cm}}
\caption{ The $g$-band QSO number counts from the final 2SLAQ survey at
  $0.4<z<2.1$ (filled circles) compared to the previous estimate from
  R05 (open circles) and the 2QZ sample (Croom et
  al. 2004; filled triangles; assuming $g=\bj-0.045$).  All samples are
  limited to $M_{\rm 
  g}<-22.5$ (or $\mbj<-22.5$).  The final 2SLAQ points (filled circles)
  have been calculated using a $z=0$ power law K-correction with
  $\alpha_\nu=-0.5$ to match R05.
  The 2QZ points use the Cristiani \& Vio (1990) K-correction which is
  very close to the $\alpha_\nu=-0.5$ version.  The new and old
  estimates from 2SLAQ agree well.  They also agree with 2QZ at
  $g<20$, but fainter than this 2QZ has significantly lower counts.
  The drop in the faintest magnitude bin for 2SLAQ is largely due
  to the $M_{\rm g}<-22.5$ limit which preferentially removes the
  faintest sources.}
\label{fig:nm2slaq2qz}
\end{figure}

\begin{figure}
\centering
\centerline{\psfig{file=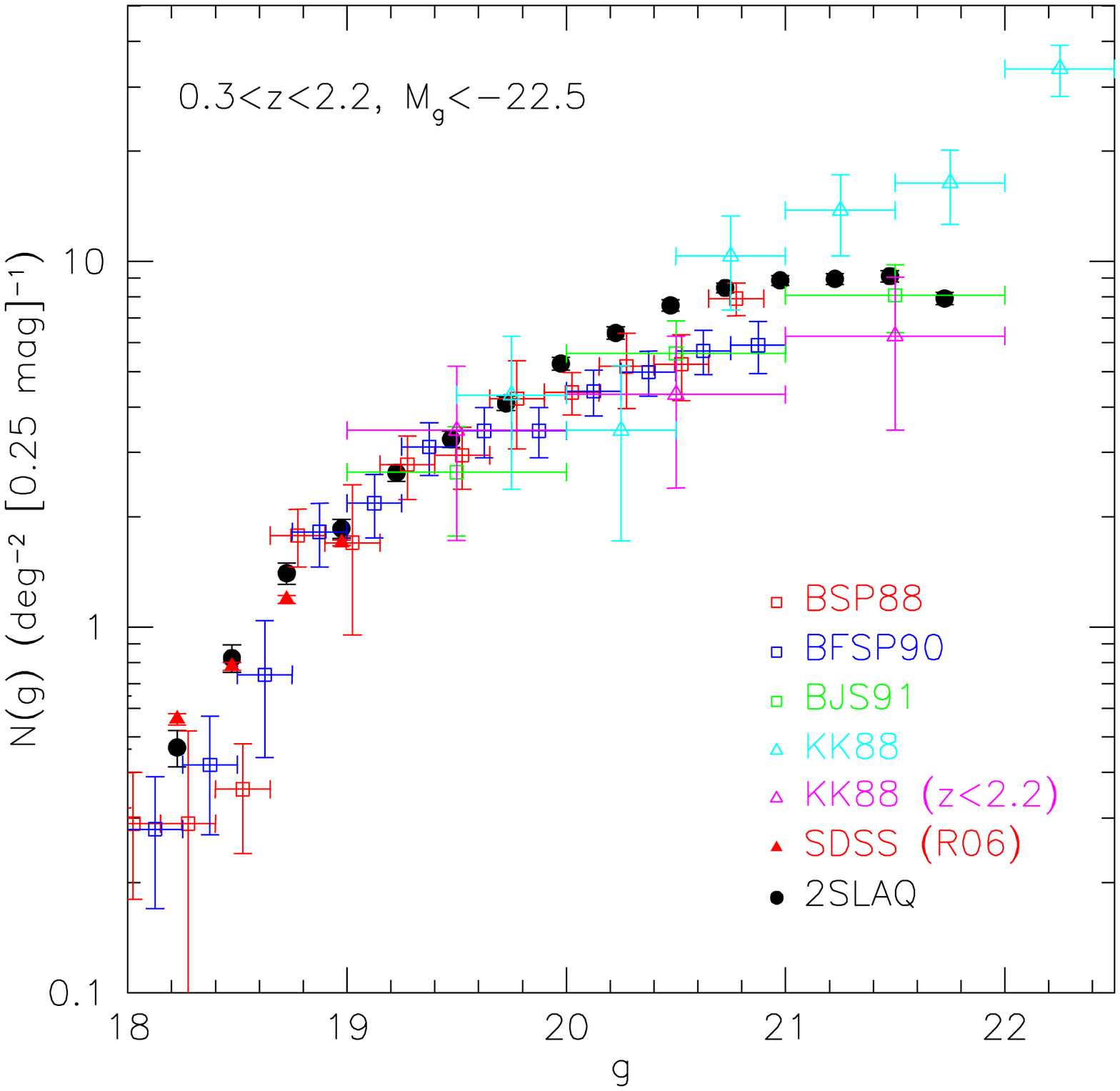,width=8cm}}
\caption{ The $g$-band QSO number counts from the final 2SLAQ survey
  (filled circles) at $0.3<z<2.2$ ($\mg<-22.5$; $\alpha_\nu=-0.5$
  K-correction) compared to other samples.  We show number
  counts from R06 (SDSS DR3) in the $g$-band
  (filled red triangles); Boyle et al. (1988; BSP88, open
  red squares); Boyle et al. (1990; BFSP90, open blue squares); Boyle et
  al. (1991; BJS91, open green squares); Koo \& Kron (1988; KK88, open cyan
  triangles); Koo \& Kron (1988) $z<2.2$, taken from the table of Boyle
  et al. (1991) (open magenta triangles).}
\label{fig:nmother}
\end{figure}

Several studies have determined K-corrections for QSOs.  Typically, to
K-correct to $z=0$, a power law correction of the form
$K(z)=-2.5(1+\alpha_\nu)\log_{10}(1+z)$ has been used, with
$\alpha_\nu\simeq-0.5$.  While QSOs have an underlying power law
continuum, the broad emission lines in their spectra have a
significant impact on the total flux in a given band.  Cristiani \&
Vio (1990), in their K-correction analysis, include the impact of
emission lines using composite QSO spectra.  Wisotzki (2000) derived
K-corrections from the optical/UV 
spectrophotometry of QSOs made available by Elvis et al. (1994).  This
study resulted in a K-correction that is substantially steeper at low
redshift ($z<0.5$) than other work, flattening to a more typical
power law at higher redshift.  R06 suggested removing
the emission line contributions to the flux before
determining a luminosity.  This gives a more direct measurement of the
energy output from the central engine, unbiased by the
location of the pass-band with respect to the QSO emission line
spectrum.  Following R06 we also correct for the emission line
flux and construct an emission line K-correction, $\kem$.  The
emission line K-correction includes a contribution from emission
lines to the $g$-band flux at $z=0$ (predominantly H$\beta$,
H$\gamma$, \oiii\ and some iron emission) and thus is not zero at
$z=0$.  To determine $\kem$ we
take a median of the emission line K-corrections derived 
from the simulated QSO spectra constructed by C09.  In
this emission line K-correction we also include the effect 
of Lyman-$\alpha$ forest absorption. 

\begin{figure}
\centering
\centerline{\psfig{file=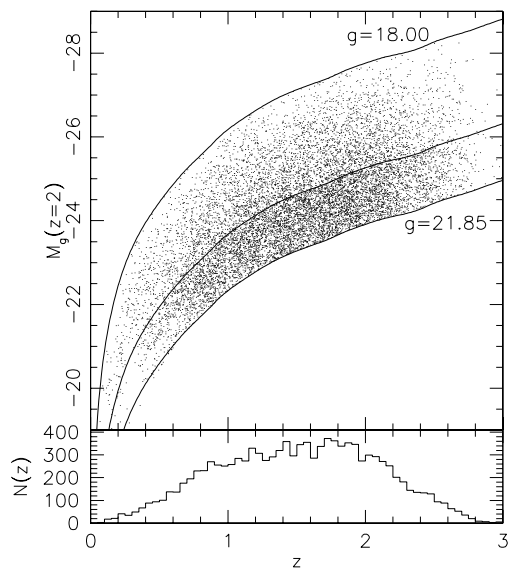,width=7.0cm}}
\caption{The $\mg(z=2)$ vs. $z$ distribution for the 2SLAQ sample,
  applying the K-correction described in Section 
  \ref{sec:qsokcorr} ($g$-band continuum, normalized at $z=2$). Each
  object is corrected for a statistical contribution from its host
  galaxy.  The top and bottom solid lines denote the
  2SLAQ apparent magnitude limits at $g=18.0$ and $g=21.85$ (not
  corrected for the host contribution).  The middle solid line
  indicates $g=20.5$ boundary between our bright and faint samples.
  The objects fainter than the flux limit at low
  redshifts are due to the host galaxy correction.}
\label{fig:mgz}
\end{figure}

It is also useful to normalize the K-correction closer to the median
redshift of the sample (e.g. Blanton et al. 2003), to minimize the
extrapolation required for the bulk of the objects.  Although the
mean redshift of the 2SLAQ sample is $z\sim1.4$, we will normalize our
K-corrections to $z=2$ in order to be consistent with R06.  This $z=2$
$g$-band K-correction acts as a $g$--band filter with the
wavelengths divided by $(1+z)$.  To obtain a total K-correction,
we then add a power law component with $\alpha_\nu=-0.3$.  This
power law slope is slightly different from the standard
$\alpha_\nu=-0.5$ usually assumed, but was found to give the best
match to observed QSO colours in the simulations of C09.

A comparison of different K-correction estimates in the literature is shown in
Fig. \ref{fig:kcorr}a.  In the cases which do not explicitly include
the $z=0$ emission line correction we have added this
contribution, so that these provide K-corrections to a continuum
magnitude at $z=0$.  The QSO K-corrections normalized at $z=0$
increasingly diverge towards high redshift.  In contrast, if we 
normalize the K-corrections at $z=2$ (Fig. \ref{fig:kcorr}b), the
K-corrections are much more consistent over the redshift range we are
sampling, $z\simeq0.4-2.6$.  We note that the 2SLAQ K-corrections and
those of Wisotzki (2000) match very well, apart from at $z=0$.
However, there is a substantial difference between the emission line-only
components of these K-corrections (Fig. \ref{fig:kcorr}c).  We
attribute this difference to the difficulty of 
defining the {\it true} continuum regions of QSO spectra, as opposed to
regions which are just free from the major emission lines, but which may
still contain significant contributions from iron lines and other
weaker emission features.

In our analysis below we will use our K-corrections (black line in
Fig. \ref{fig:kcorr}), and correct to the continuum flux in the
$g$-band at $z=2$.  The emission line contribution at this
redshift is $0.194$ mag, such that $M_{\rm g,cont}(z=2)=M_{\rm
  g,total}(z=2)+0.194$.  R05 used a more traditional K-correction,
zero-pointed at $z=0$, assuming a power law with $\alpha_{\nu}=-0.5$
and no emission line correction.  Averaged over the sample at
$z=0.3-2.2$ we find that we need to subtract 0.41 from the R05
absolute magnitudes to match our new K-corrections (with a maximum
variation with redshift of $\pm0.1$ mag).  We find that an
identical correction of $-0.41$ is needed to transform the Cristiani
\& Vio (1990) K-corrections onto our new K-corrections.  In some
cases below, we will be required to use the different K-corrections
discussed above to compare our results to those of previous authors.

\section{QSO number counts}\label{sec:nm}

\begin{figure*}
\centering
\centerline{\psfig{file=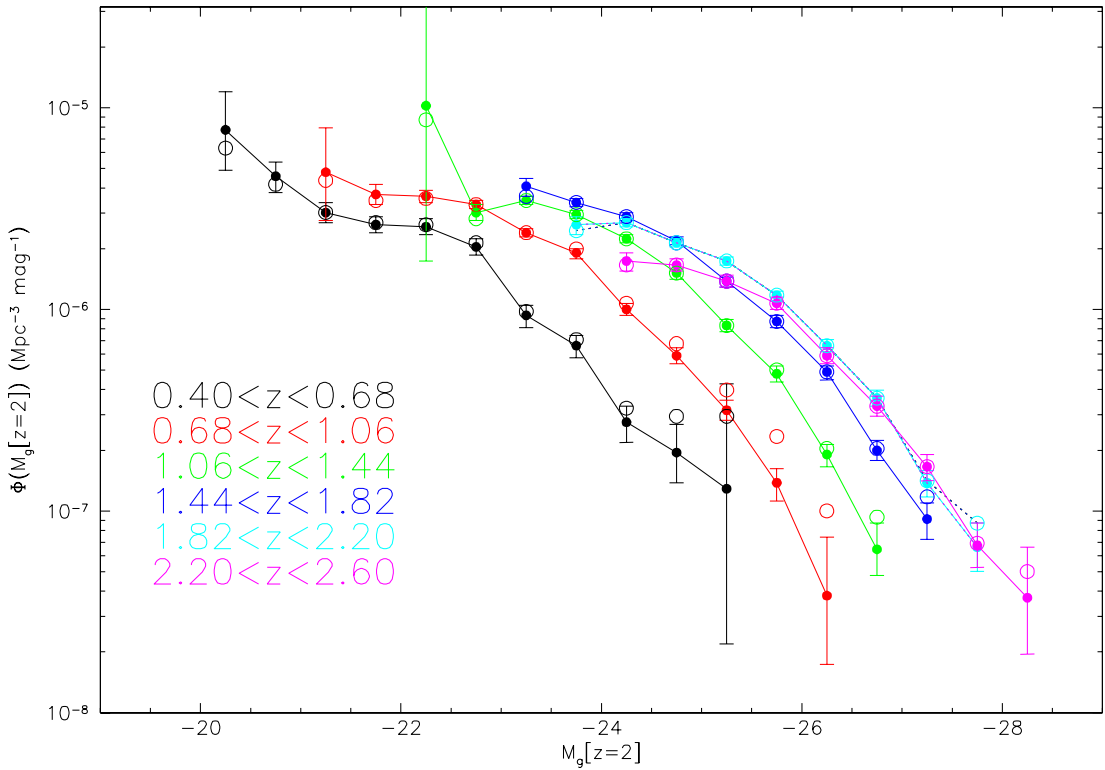,width=16cm,angle=270}}
\caption{The binned 2SLAQ luminosity function for 6 redshift intervals
  from $z=0.4$ to $z=2.6$.  The filled points are those derived using
  the model weighted estimator described in the text.  The open points
  are derived using the Page \& Carrera (2000) estimator which
  accounts for the flux limits crossing the bins, but does not account
  for the change in source density within a bin.}
\label{fig:lfmg}
\end{figure*}

We first calculate the number counts of QSOs as a function of $g$
magnitude.  The raw QSO counts from the 2SLAQ sample are shown in
Fig. \ref{fig:nmcomp} (open squares).  The filled squares show
the number counts corrected for coverage completeness only; that is
fraction of targets in our survey region which where actually observed.
The coverage completeness is known exactly, while other completeness
corrections (e.g. photometric selection, spectroscopic completeness;
see C09 for details) have some uncertainties. The fully corrected
counts (in the redshift range $0.4<z<2.1$) are shown by the open and
filled circles (for the NGP and SGP regions respectively).  We see
no significant difference between the NGP and SGP regions.
Fig. \ref{fig:nm2slaq2qz} compares the number counts from the
preliminary 2SLAQ sample (R05) to our current determination.  In order
to directly compare the two, we use the same K-correction and $\mg$
limit as R05, namely, a power law K-correction, normalized at $z=0$
with $\alpha_\nu=-0.5$ and a limit of $M_{\rm g}<-22.5$.  We also do
not make any correction for the host galaxy contribution to the total
flux, as this was not done by R05.  Despite the different photometric
completeness estimates used in our work 
and R05, there is excellent agreement between the two
estimates of the 2SLAQ $n(g)$ distribution.  In particular, the
decline at the faintest magnitudes is present in both analyses.  This
drop is mostly due to the absolute magnitude cut which preferentially
removes objects at the faintest magnitudes.

In Fig. \ref{fig:nm2slaq2qz} we also compare the 2SLAQ $n(g)$ to that
derived from the 2QZ sample (Croom et al. 2004).  As noted by R05, the
2SLAQ counts are an excellent match to 2QZ at $g<20$, but at fainter
magnitudes 2SLAQ contains an increasingly higher density of QSOs than
does 2QZ, with this excess reaching $\sim25$ per cent at the faintest bin
of the 2QZ.  We discuss this discrepancy between 2SLAQ and 2QZ number
counts in further detail in Section \ref{sec:disnm}.   In
Fig. \ref{fig:nmother} we compare the  
2SLAQ number counts (again using the $z=0$, $\alpha_\nu=-0.5$ power
law K-correction and $\mg<-22.5$) to a wide range of
previously published number counts.  This includes the main SDSS
sample (R06, filled red triangles), which is only
plotted at $g<19$, after which incompleteness strongly effects the
$g$-band number counts (as the sample is selected in the
$i$-band).  We also show the older number counts of Boyle et al. (1988;
1990).  These have a similar depth to 2QZ and are also consistently
below 2SLAQ at $g>20$.  The deeper sample of Boyle, Jones \& Shanks
(1991) is in agreement with 2SLAQ (albeit with large errors), while
the full Koo \& Kron (1988) sample, not limited to $z<2.2$, lies above
2SLAQ.  As pointed out by R05, 
the number counts for the sub-sample of Koo \& Kron (1988) QSOs
limited to $z<2.2$ is significantly lower, and is below or consistent
with 2SLAQ (although the KK88 sample
is also restricted to $z>0.9$).  Overall there is good agreement
between these various analyses; however, the redshift ranges
and limits of these different samples do not always match exactly.
Detailed comparison to more recent estimates of the QSO luminosity
function will be presented below.  
   
\section{The luminosity function}\label{sec:binnedlf}

\begin{figure*}
\centering
\centerline{\psfig{file=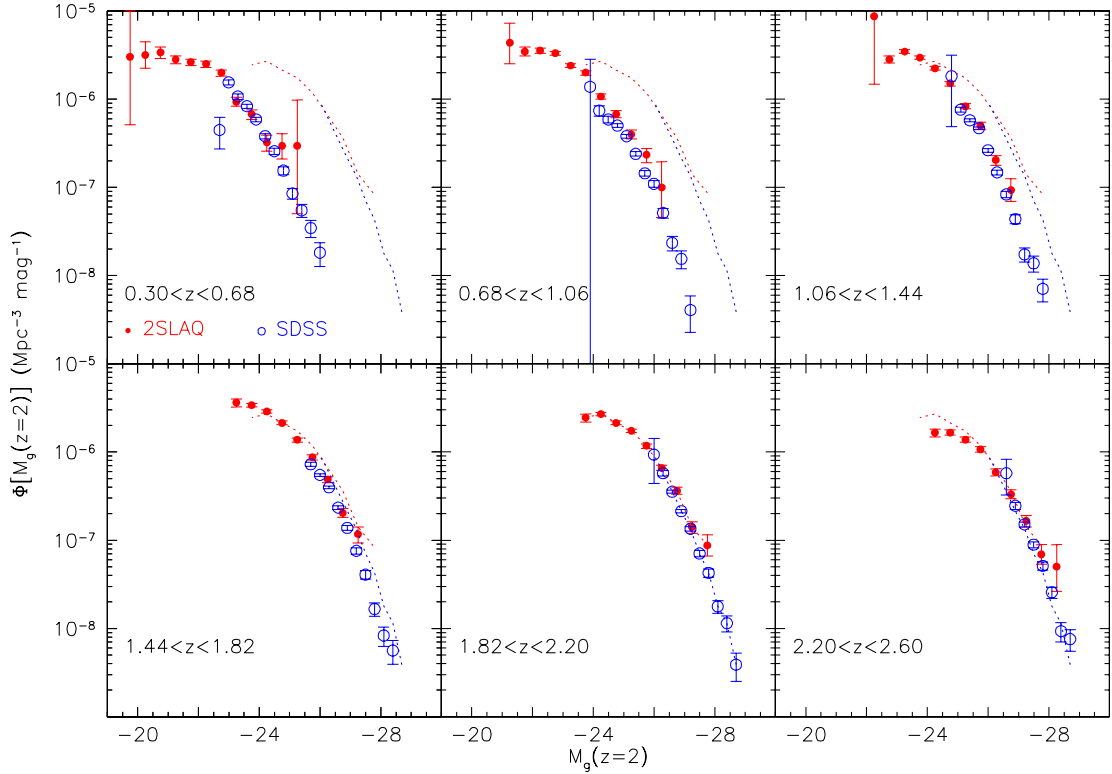,width=16cm,angle=270}}
\vspace{0.5cm}
\caption{The binned 2SLAQ luminosity function (filled red points) for
  six redshift intervals from $z=0.3$ to $z=2.6$, compared to the SDSS
  LF (Richards et al. 2006; blue open points).  The dotted lines show
  the LFs at $1.82<z<2.20$ as a reference.}
\label{fig:lfsdss}
\end{figure*}

\begin{figure}
\centering
\centerline{\psfig{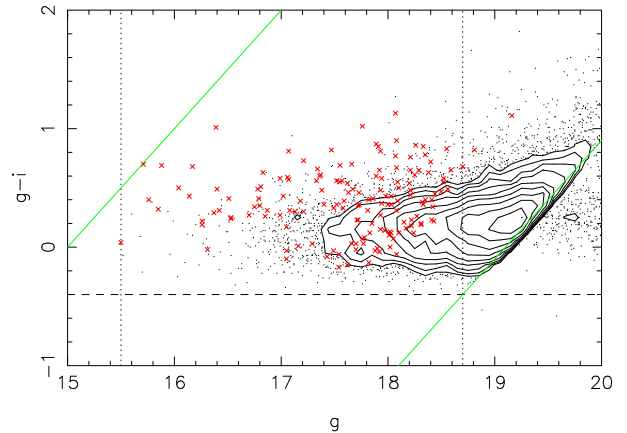}}
\caption{The $g-i$ colours vs. $g$ of SDSS QSOs from the DR3 LF sample
  of Richards et al (2006).  The black points and contours
  are SDSS QSOs at $z>0.3$, while the red crosses are at $z<0.3$.
   The contours are logarithmically spaced, based on the density of
  points per 0.1 by 0.1 mag bin, starting at log(density)=1.0, with
  steps of 0.17.  The
  diagonal solid (green) lines mark the SDSS flux limits at $i=15.0$ and
  $i=19.1$.  The horizontal dashed line marks the limit of $g-i=-0.4$,
  below which there are virtually no QSOs.  The vertical dotted lines
  indicate our chosen $g$-band limits of $g=15.5$ and $g=18.7$.}
\label{fig:sdssgig}
\end{figure}

\begin{figure*}
\centering
\centerline{\psfig{file=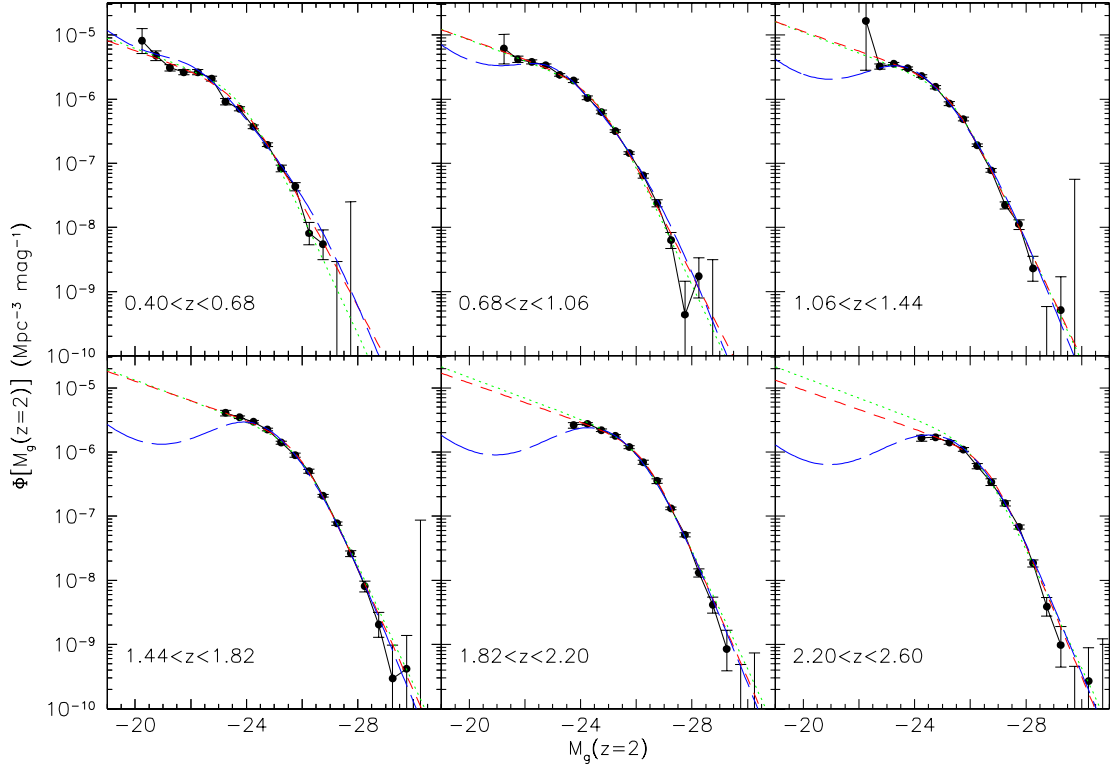,width=16cm,angle=270}}
\vspace{3mm}
\caption{The combined 2SLAQ and SDSS $g$-band luminosity function for
  six redshift intervals from $z=0.4$ to $z=2.6$, estimated using the
  model-weighted method.  The error bars without visible points are
  upper limits, i.e. no QSOs were found, although the accessible
  volume was non-zero.  We compare the measured LF to the best PLE model
  (green dotted lines), smooth LDDE model (blue long--dashed lines)
  and LEDE model (red short--dashed lines).} 
\label{fig:lfcomb}
\end{figure*}

In Fig. \ref{fig:mgz} we show the distribution of 2SLAQ sources in the
$z-\mg$ plane, applying the $z=2$ $g$-band continuum K-correction
described in Section \ref{sec:qsokcorr}.  When we use $\mg$ hereafter
we will be using this as shorthand for continuum $\mg(z=2)$. 
Using this sample we first
calculate the binned luminosity function using the model weighted
estimator suggested by Miyaji, Hasinger \& Schmidt (2001).  This
improves on the $1/V$ estimator devised by Page \& Carrera (2000),
which partially corrects for binning effects, but assumes a uniform
distribution across each bin.  The model weighted estimator uses the
best model fit to the {\it unbinned} LF data (described in Section
\ref{sec:mlfits}) to correct for the variation of the LF within  
a bin, which is particularly critical at the steep bright end of the
QSO LF.  This estimator gives the binned LF as
\begin{equation}
\Phi(M_{\rm g,i},z_{\rm i})=\Phi(M_{\rm g,i},z_{\rm i})^{\rm
  model}\frac{N^{\rm obs}_{\rm i}}{N^{\rm model}_{\rm i}},
\end{equation}
where $M_{\rm g,i}$ and $z_{\rm i}$ are the absolute magnitude and redshift at
the centre of the $i$th bin. $\Phi(M_{\rm g,i},z_{\rm i})^{\rm model}$
is the best fit model evaluated at $M_{\rm g,i}$ and $z_{\rm i}$.  $N^{\rm
  obs}_{\rm i}$ is the number of observed QSOs in the bin and $N^{\rm
  model}_{\rm i}$ the number estimated from the model (accounting for
completeness corrections).  In the luminosity functions presented here
we use the luminosity evolution + density evolution (LEDE) model fits
described in Section \ref{sec:ledefits},
although assuming a PLE model fit does not produce a
significant difference to the estimate of the binned LF.

Fig. \ref{fig:lfmg} shows the binned 2SLAQ LF for six different
redshift intervals from $z=0.4$ to $z=2.6$.  This binning matches
the redshift intervals from the SDSS DR3 luminosity function of
R06, except that we have increased the lowest redshift
limit from $0.3$ to $0.4$, which is the lowest redshift at which we
trust our completeness corrections.  The 2SLAQ LF shows the approximate
characteristics of PLE, that is, a fixed LF
shape with an evolving $L^*$.  However, there is also evidence for
deviations from a PLE model, including variation in slope and 
normalization which we will investigate below.  The open points
in Fig. \ref{fig:lfmg} show the binned LF using the $1/V$ Page \&
Carrera (2000) estimator.  This demonstrates that at the steep bright
end of the QSO LF, the $1/V$ estimator can cause significant bias.

\subsection{Comparison to the SDSS LF and a combined 2SLAQ-SDSS LF}

\begin{table*}
\caption{Binned luminosity function for the combined 2SLAQ and SDSS
  sample using the model weighted estimator, as plotted in
  Fig. \ref{fig:lfcomb}.  We give the value of $\lp$ in 6 redshift
  intervals, and in $\Delta\mg=0.5$ mag 
bins.  We also list the mean redshift ($\bar{z}$) in each bin, the
number of QSOs contributing to the LF ($\nq$) and the lower and upper
errors ($\dlp$).  At the bright end of the LF some bins contain no
  QSOs, even though the accessible volume is non-zero.  In this case
  the values in the $\lp$ column are $1\sigma$ Poisson upper limits on
  the LF in this bin.} 
\label{tab:binnedlf}
\begin{tabular}{@{}ccrccccrccccrccc@{}}
\hline
&\multicolumn{5}{c}{$0.40<z<0.68$}&\multicolumn{5}{c}{$0.68<z<1.06$}&\multicolumn{5}{c}{$1.06<z<1.44$}\\
$\mg$ &$\bar{z}$&$\nq$&$\lp$&\multicolumn{2}{c}{$\dlp$}&$\bar{z}$ &$\nq$&$\lp$ &\multicolumn{2}{c}{$\dlp$}&$\bar{z}$ &$\nq$&$\lp$ &\multicolumn{2}{c}{$\dlp$} \\
\hline 
--29.75 &  --  &    0 &   --   &   --   &  --   &  --  &    0 &   --   &   --   &  --   &  --  &    0 & --7.25 &   --   &  --  \\
--29.25 &  --  &    0 &   --   &   --   &  --   &  --  &    0 &   --   &   --   &  --   & 1.31 &    1 & --9.31 & --0.77 & +0.52\\
--28.75 &  --  &    0 &   --   &   --   &  --   &  --  &    0 & --8.50 &   --   &  --   &  --  &    0 & --9.23 &   --   &  --  \\
--28.25 &  --  &    0 &   --   &   --   &  --   & 1.02 &    3 & --8.85 & --0.34 & +0.29 & 1.25 &    7 & --8.65 & --0.20 & +0.19\\
--27.75 &  --  &    0 & --7.60 &   --   &  --   & 0.77 &    1 & --9.42 & --0.77 & +0.52 & 1.30 &   34 & --7.96 & --0.08 & +0.07\\
--27.25 &  --  &    0 & --8.53 &   --   &  --   & 0.97 &   15 & --8.25 & --0.13 & +0.12 & 1.29 &   69 & --7.65 & --0.06 & +0.05\\
--26.75 & 0.56 &    5 & --8.28 & --0.24 & +0.22 & 0.94 &   57 & --7.67 & --0.06 & +0.05 & 1.30 &  254 & --7.11 & --0.03 & +0.03\\
--26.25 & 0.60 &    8 & --8.10 & --0.18 & +0.17 & 0.92 &  158 & --7.24 & --0.04 & +0.03 & 1.24 &  431 & --6.73 & --0.02 & +0.02\\
--25.75 & 0.58 &   43 & --7.37 & --0.07 & +0.06 & 0.92 &  354 & --6.88 & --0.02 & +0.02 & 1.20 &  260 & --6.32 & --0.03 & +0.03\\
--25.25 & 0.57 &   81 & --7.09 & --0.05 & +0.05 & 0.85 &  369 & --6.52 & --0.02 & +0.02 & 1.27 &  194 & --6.08 & --0.03 & +0.03\\
--24.75 & 0.58 &  200 & --6.71 & --0.03 & +0.03 & 0.83 &  262 & --6.22 & --0.03 & +0.03 & 1.28 &  300 & --5.82 & --0.03 & +0.02\\
--24.25 & 0.54 &  250 & --6.44 & --0.03 & +0.03 & 0.90 &  173 & --6.00 & --0.03 & +0.03 & 1.27 &  448 & --5.65 & --0.02 & +0.02\\
--23.75 & 0.50 &  185 & --6.16 & --0.03 & +0.03 & 0.90 &  287 & --5.72 & --0.03 & +0.02 & 1.26 &  555 & --5.53 & --0.02 & +0.02\\
--23.25 & 0.54 &   77 & --6.04 & --0.05 & +0.05 & 0.90 &  342 & --5.62 & --0.02 & +0.02 & 1.24 &  484 & --5.46 & --0.02 & +0.02\\
--22.75 & 0.56 &  121 & --5.69 & --0.04 & +0.04 & 0.87 &  421 & --5.48 & --0.02 & +0.02 & 1.14 &  118 & --5.52 & --0.04 & +0.04\\
--22.25 & 0.58 &  141 & --5.59 & --0.04 & +0.04 & 0.83 &  274 & --5.44 & --0.03 & +0.03 & 1.07 &    1 & --4.99 & --0.77 & +0.52\\
--21.75 & 0.55 &  119 & --5.58 & --0.04 & +0.04 & 0.75 &   81 & --5.43 & --0.05 & +0.05 &  --  &    0 &   --   &   --   &  --  \\
--21.25 & 0.52 &   81 & --5.52 & --0.05 & +0.05 & 0.70 &    5 & --5.32 & --0.24 & +0.22 &  --  &    0 &   --   &   --   &  --  \\
--20.75 & 0.49 &   33 & --5.34 & --0.08 & +0.07 &  --  &    0 &   --   &   --   &  --   &  --  &    0 &   --   &   --   &  --  \\
--20.25 & 0.45 &    7 & --5.11 & --0.20 & +0.19 &  --  &    0 &   --   &   --   &  --   &  --  &    0 &   --   &   --   &  --  \\
\hline
&\multicolumn{5}{c}{$1.44<z<0.82$}&\multicolumn{5}{c}{$1.82<z<2.20$}&\multicolumn{5}{c}{$2.20<z<2.60$}\\
$\mg$ &$\bar{z}$&$\nq$&$\lp$&\multicolumn{2}{c}{$\dlp$}&$\bar{z}$ &$\nq$&$\lp$ &\multicolumn{2}{c}{$\dlp$}&$\bar{z}$ &$\nq$&$\lp$ &\multicolumn{2}{c}{$\dlp$} \\
\hline
--30.25 &  --  &    0 &   --   &   --   &  --   &  --  &    0 &   --   &   --   &  --   &  --  &    0 & --9.81 & --     &  --  \\
--30.25 &  --  &    0 & --7.06 &   --   &  --   &  --  &    0 & --9.13 &   --   &  --   & 2.21 &    1 & --9.58 & --0.77 & +0.52\\
--29.75 & 1.76 &    1 & --9.40 & --0.77 & +0.52 &  --  &    0 & --9.31 &   --   &  --   &  --  &    0 & --9.34 &   --   &  --  \\
--29.25 & 1.48 &    1 & --9.54 & --0.77 & +0.52 & 2.03 &    3 & --9.08 & --0.34 & +0.29 & 2.51 &    3 & --9.02 & --0.34 & +0.29\\
--28.75 & 1.67 &    7 & --8.69 & --0.20 & +0.19 & 2.03 &   15 & --8.38 & --0.13 & +0.12 & 2.45 &   12 & --8.42 & --0.15 & +0.14\\
--28.25 & 1.65 &   28 & --8.09 & --0.09 & +0.08 & 2.01 &   47 & --7.89 & --0.07 & +0.06 & 2.38 &   61 & --7.74 & --0.06 & +0.05\\
--27.75 & 1.66 &   90 & --7.58 & --0.05 & +0.04 & 2.02 &  193 & --7.30 & --0.03 & +0.03 & 2.37 &  212 & --7.17 & --0.03 & +0.03\\
--27.25 & 1.66 &  285 & --7.11 & --0.03 & +0.02 & 1.99 &  387 & --6.88 & --0.02 & +0.02 & 2.33 &  108 & --6.81 & --0.04 & +0.04\\
--26.75 & 1.61 &  541 & --6.68 & --0.02 & +0.02 & 1.99 &  150 & --6.47 & --0.04 & +0.03 & 2.38 &   73 & --6.48 & --0.05 & +0.05\\
--26.25 & 1.61 &  195 & --6.32 & --0.03 & +0.03 & 2.00 &  185 & --6.18 & --0.03 & +0.03 & 2.36 &  119 & --6.23 & --0.04 & +0.04\\
--25.75 & 1.65 &  226 & --6.06 & --0.03 & +0.03 & 2.01 &  285 & --5.93 & --0.03 & +0.02 & 2.39 &  201 & --5.97 & --0.03 & +0.03\\
--25.25 & 1.64 &  313 & --5.86 & --0.03 & +0.02 & 2.01 &  432 & --5.76 & --0.02 & +0.02 & 2.37 &  277 & --5.86 & --0.03 & +0.03\\
--24.75 & 1.65 &  510 & --5.66 & --0.02 & +0.02 & 2.00 &  538 & --5.67 & --0.02 & +0.02 & 2.38 &  265 & --5.78 & --0.03 & +0.03\\
--24.25 & 1.63 &  635 & --5.54 & --0.02 & +0.02 & 1.99 &  557 & --5.57 & --0.02 & +0.02 & 2.31 &  103 & --5.76 & --0.05 & +0.04\\
--23.75 & 1.63 &  528 & --5.47 & --0.02 & +0.02 & 1.91 &  101 & --5.58 & --0.05 & +0.04 &  --  &    0 &   --   &   --   &  --  \\
--23.25 & 1.52 &   98 & --5.39 & --0.05 & +0.04 &  --  &    0 &   --   &   --   &  --   &  --  &    0 &   --   &   --   &  --  \\
\hline
\end{tabular}
\end{table*}

\begin{figure}
\centering
\centerline{\psfig{file=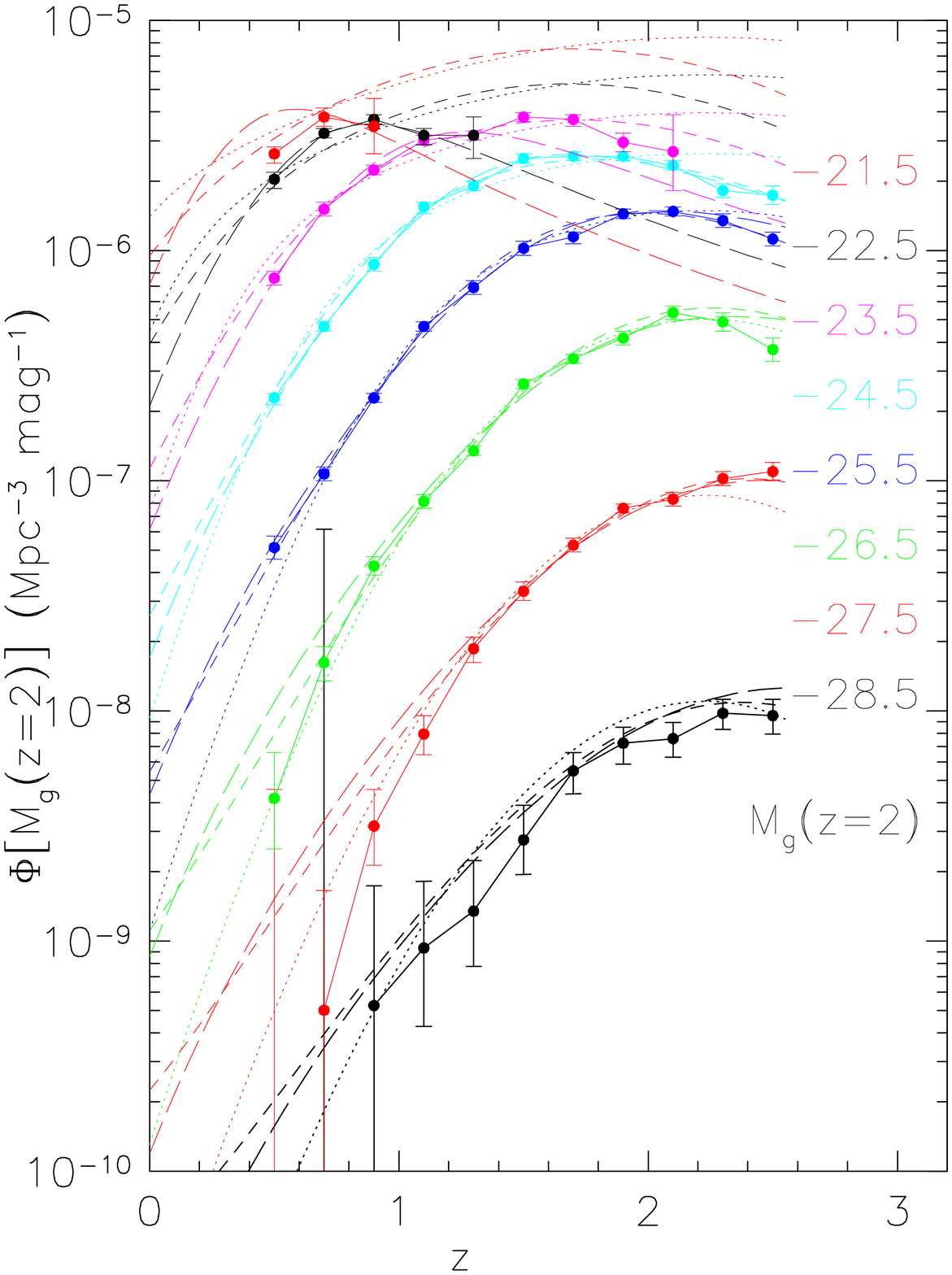,width=9cm}}
\caption{The combined 2SLAQ and SDSS luminosity function (using the
  model weighted method) plotted as a function of redshift for
  different $\mg$ intervals.  The brightest intervals are at the
  bottom of the plot and the faintest at the top.  Each $\mg$ interval
  is connected by a solid line apart from points which are upper
  limits (e.g. $\mg=-27.5$, $z=0.5$).  To aid the clarity of this
  plot, the small number of noisy points at $\mg=-29.5$ and
  $\mg=-30.5$ have been omitted.  We compare the measured LF to the
  best fit PLE model (dotted lines), smooth LDDE model (long dashed
  lines) and LEDE model (short dashed lines).} 
\label{fig:lfz}
\end{figure}

In Fig. \ref{fig:lfsdss} we compare the
2SLAQ LF to that derived for the brighter SDSS DR3 sample by R06.  As
both analyses K-correct to a continuum--only magnitude, we can convert
from the $\mi(z=2)$ magnitudes of R06 with 
\begin{equation}
\mg(z=2)=\mi(z=2)+2.5\alpha_\nu\log\left(\frac{4670{\rm \AA}}{7471{\rm
    \AA}}\right).
\label{eq:mgmi}
\end{equation}
In Fig. \ref{fig:lfsdss} we assume $\alpha_\nu=-0.5$.  We
plot the LF in six different panels, and the dotted lines show the
$1.82<z<2.20$ LF as a comparison.  The SDSS DR3 LF
(open circles) is a smooth continuation of the 2SLAQ LF towards higher
luminosities.  This figure clearly shows the QSO LF break.  This break
is a gradual flattening of the 
QSO LF, which starts at approximately the faint limit of the main SDSS
QSO sample.

We next combine the 2SLAQ and SDSS data sets to produce a single, binned,
$g$-band luminosity function with unprecedented precision and dynamic
range.  This is done by taking the QSO sample presented by R06,
including the completeness estimates.  The R06 LF is presented in the
$i$-band, as this is the band which provides the flux limit for the SDSS
QSO sample.  To combine this data with the 2SLAQ sample we need to
convert the selection function (Table 1 of R06) from the $i$-band to
the $g$-band, and we need to specify an appropriate range in $g$-band
flux over which to calculate the luminosity function.  To convert the
selection function to the $g$-band we calculate the median $g-i$
colours of SDSS QSOs as a function of redshift, $z$.  This is then
used to map the completeness in each ($i$-band, $z$) interval to the
corresponding ($g$-band, $z$) interval.  The ($g$-band, $z$) completeness is then
re-sampled onto a uniform grid.  To obtain suitable flux limits for
the SDSS sample in the $g$-band we examine the $g-i$ colours of SDSS
QSOs as a function of $g$ (for all redshifts; Fig. \ref{fig:sdssgig}).
The $i$-band flux limits imposed on the SDSS QSO sample (solid
diagonal lines in  Fig. \ref{fig:sdssgig}) mean that at faint $g$-band
magnitudes the bluest QSOs will be missed.  We apply a faint limit of
$g=18.7$, which does not cause any QSOs to be rejected on the basis of
their $g-i$ colour, as there are virtually no QSOs bluer than
$g-i=-0.4$ in the SDSS sample.
At the bright end, there are no bright $z>0.3$ QSOs
($g~\simlt~17$) which are redder than $g-i=0.5$.  Those at $z<0.3$ are
systematically redder (red crosses in Fig. \ref{fig:sdssgig}), due to
the contribution of their host galaxy.  However, we are not
considering the QSO LF below $z=0.3$, therefore we can safely ignore
this population.  We therefore set the bright limit at $g=15.5$
(cf. the SDSS $i$-band bright limit of $i=15.0$).

Applying the appropriate flux limits and completeness corrections, we
then combine the SDSS and 2SLAQ samples to produce the binned luminosity
function in Fig. \ref{fig:lfcomb} (using the model weighted
estimator).  This LF covers a much greater dynamic
range that either the SDSS or 2SLAQ samples do on their own.  One
point to note is that the binning has a non-trivial effect on the LF
of the combined data set, near the overlap
of the two samples.  The model weighted LF estimator accurately
corrects for these binning biases, and it is this estimator that is 
shown in Fig. \ref{fig:lfcomb}.  The LF values and associated errors
are listed in Table \ref{tab:binnedlf}.  

In Fig. \ref{fig:lfz} we plot the space density of QSOs from the
combined 2SLAQ+SDSS sample as a function of redshift in narrow $\mg$
slices.  At bright absolute magnitudes ($\mg~\simlt~-27$) the space
density of QSOs is monotonically increasing up to $z\simeq2.5$.
However, the space density of fainter QSOs peaks at lower
redshift, e.g. $z\simeq1.6$ for $\mg=-23.5$.  This is in
disagreement with PLE, in which the space density of QSOs
peaks at the same redshift at every luminosity.  This is the same
trend that has been seen in previous X-ray selected samples of AGN and
has been called ``AGN downsizing'' (e.g. Barger et al. 2005; Hasinger
et al. 2005).  This downsizing was first seen in X-ray samples because
they were the first to have the dynamic range and object numbers to
allow it to be measured.

\subsection{Comparison to other observed LFs}

\begin{figure*}
\centering
\centerline{\psfig{file=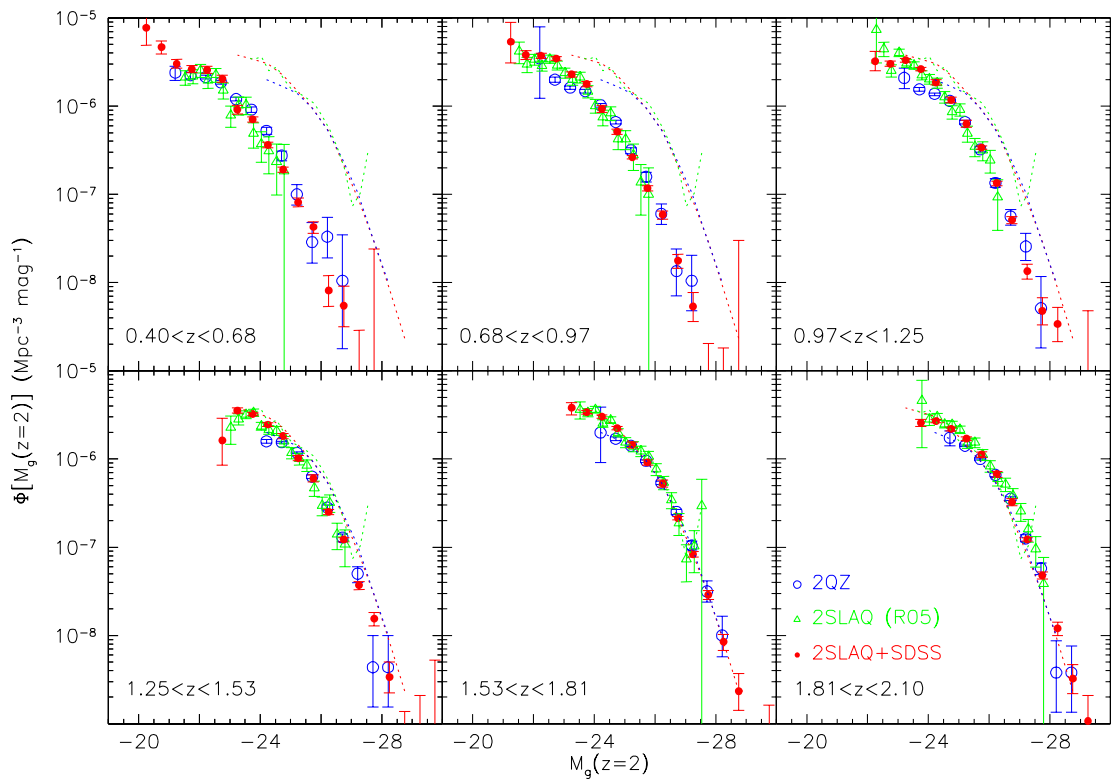,width=16cm,angle=270}}
\vspace{0.5cm}
\caption{The binned 2SLAQ+SDSS luminosity function for six redshift intervals
  from $z=0.4$ to $z=2.1$ (filled red points), compared to the 2QZ LF
  of C04 (open blue circles) and the preliminary 2SLAQ LF of R05
  (open green triangles).  The dotted lines show the LFs at
  $1.53<z<1.81$ as a reference.}
\label{fig:lf2qz}
\end{figure*}

Fig. \ref{fig:lf2qz} presents a comparison of our combined 2SLAQ and SDSS
LF to that measured from the 2QZ survey by C04.
To convert from $\bj$ to $g$-band magnitude we use $g-\bj=-0.045$ as
found by R05.  We then take into account the difference between
our current K-correction and that used by C04, giving $\mg(z=2)=M_{\rm
  b_{\rm J}}-0.455$ (averaged over redshift).  In general there is 
excellent agreement between the luminosity functions measured from the
two samples.  In the lowest redshift interval ($0.4<z<0.68$) the 2QZ
LF appears slightly higher than the 2SLAQ data at $\mg\simeq-24$,
which we attribute to the fact that the K-correction used by C04
(derived from the work of Cristiani \& Vio 1990) is flatter at low
redshift than our assumed K-correction (see Fig. \ref{fig:kcorr}).  A
shift of only $\sim0.1$ mags is sufficient to bring the 2QZ and 2SLAQ
data into excellent agreement here.  In
all other redshift intervals the agreement between 2QZ and 2SLAQ is
very good, with the exception of the faintest $\sim1$ mag of the 2QZ
LF, which is significantly lower than the 2SLAQ LF at all redshifts.
This is consistent with the difference seen in the number counts
presented in Section \ref{sec:nm} above.  In Fig. \ref{fig:lf2qz} we
also plot the preliminary 2SLAQ LF calculated by R05 (green
triangles).  We offset this LF by $-0.41$ mags to account for the
difference between our K-correction and the $z=0$, $\alpha_\nu=-0.5$
power law K-correction used by R05.  The R05 2SLAQ LF and our current
work are in excellent agreement, even though the completeness
estimates have been significantly revised, and different K-corrections
are assumed.  This gives us confidence that any remaining
uncertainties in our photometric completeness or effects due to the
K-correction are unlikely to significantly bias our LF estimates.

\begin{figure}
\centering
\centerline{\psfig{file=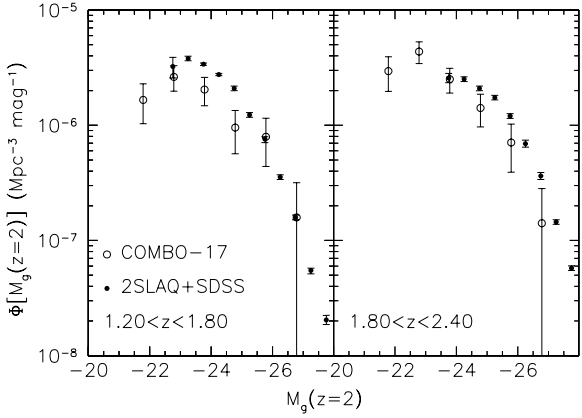,width=8cm,angle=270}}
\caption{The binned 2SLAQ+SDSS luminosity function (filled circles) compared to
  Wolf et al. (2003) (COMBO-17; open circles). }
\label{fig:lfcombo}
\end{figure}

\begin{figure}
\centering
\centerline{\psfig{file=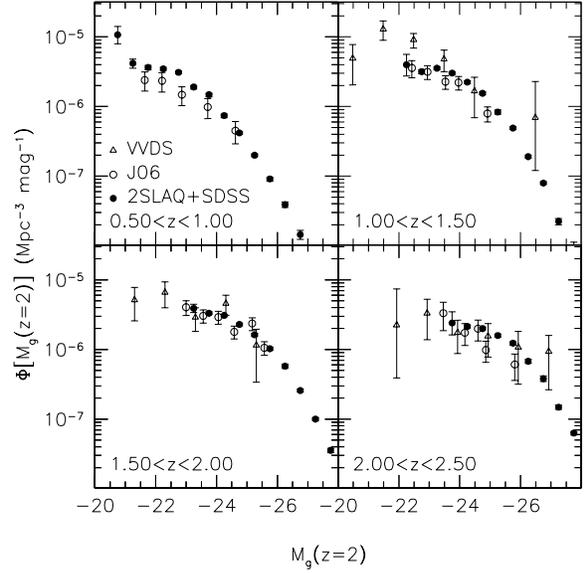,width=8cm}}
\caption{The binned 2SLAQ+SDSS luminosity function (filled circles) compared to
  the LFs of Jiang et al (2006) (open circles) and Bongiorno et al. 2007
  (VVDS, open triangles).  Note that the VVDS LF is not plotted for
  $0.5<z<1.0$, as Bongiorno et al.\ did not calculate it in this
  redshift interval.}
\label{fig:lfvvdsjiang}
\end{figure}

In Figs. \ref{fig:lfcombo} and \ref{fig:lfvvdsjiang} we compare the
2SLAQ+SDSS result to other LF
measurements at faint fluxes.  While these reach fainter fluxes than
2SLAQ, they are constructed from smaller samples and so have much larger
uncertainties.  The COMBO-17 LF \cite{combo17lf} is shown in 
Fig. \ref{fig:lfcombo}, with the 2SLAQ+SDSS LF re-binned to the same
redshift intervals.  The COMBO-17 LF is defined in the $M_{1450}$
pass-band using Vega magnitudes, which is close to the $g$-band at
$z=2$.  The equivalent of Eq. \ref{eq:mgmi} then gives
$\mg(z=2)=M_{1450}+1.216$ for $\alpha_\nu=-0.5$, after correcting from
Vega to AB \cite{og83} magnitudes.
Applying this correction gives the COMBO-17 LF plotted in Fig. 
\ref{fig:lfcombo}.  We see good agreement between 2SLAQ and COMBO-17
where they overlap.  The COMBO-17 LF is slightly lower than the 2SLAQ
LF in the $1.20<z<1.80$ interval.  However, given the size of the
errorbars and possible remaining uncertainty in the flux transformation,
we do not consider this significant.

Jiang et al. (2006) determine the QSO LF using a sample of 414 QSOs
covering an area of 3.9 deg$^2$ limited to $g<22.5$.  This sample was
selected from co-adds of the multi-epoch Stripe 82 SDSS data.  Jiang
et al. clearly demonstrate the flattening of the QSO LF towards
faint magnitudes, but do not have sufficient precision to measure
any downsizing effect.  Jiang et al. use a K-correction which is
derived separately for each object, but that is zero-pointed to
$z=0$.  To correct the Jiang et al. magnitudes we assume that they
are well approximated by a simple $\alpha_\nu=-0.5$ power law
K-correction and apply an offset of $-0.41$ to move them to our
$\mg(z=2)$ system.  The comparison of our 2SLAQ+SDSS LF with the
result of Jiang et al. is shown in Fig. \ref{fig:lfvvdsjiang}.  There
is good agreement between the two data sets, although at $0.5<z<1.0$
the Jiang et al. LF is somewhat lower than the 2SLAQ+SDSS LF.  

Another recent determination of the faint end of the QSO LF is from
the VLT VIMOS Deep Survey (VVDS; Bongiorno et al 2007).  A comparison between
2SLAQ and VVDS is shown in Fig. \ref{fig:lfvvdsjiang}.  The VVDS LF is
determined in the $B$-band.  To convert to our $\mg(z=2)$ band we use
\begin{equation}
\mg(z=2)=M_{\rm B}(z=0)-2.5(1+\alpha_\nu)\log(3)-0.14+0.209,
\end{equation}
where the $-0.14$ arises from the mean difference between the $B$ and
$g$-bands for QSOs \cite{sdssdr3lf} and the $+0.209$ comes from the
emission line contribution.  This results in a correction of
$\mg(z=2)=M_{\rm B}(z=0)-0.527$ for $\alpha_\nu=-0.5$.  The 2SLAQ and
VVDS LFs are again in good general agreement, except for the faintest
two 2SLAQ bins in the $1.0<z<1.5$ redshift interval, where the VVDS
point is significantly higher than 2SLAQ.  The 2SLAQ points in this
region are marginally lower than those at brighter luminosities,
possibly suggesting some unaccounted-for incompleteness.  However,
these 2SLAQ points are in good agreement with the result of Jiang et
al. (2006).  Given the the uncertainties due to small numbers in the
VVDS sample and issues such as cosmic variance, it is not clear that
there is a true disagreement.  Future comparison with the final VVDS
sample, covering a larger area, should aid in the resolution of this
issue.

\begin{figure}
\centering
\centerline{\psfig{file=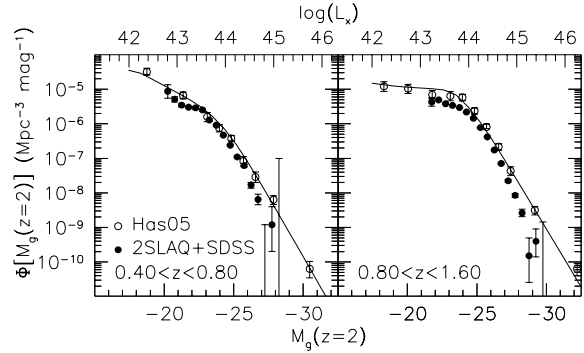,width=8cm}}
\caption{The binned 2SLAQ luminosity function (filled points) compared
  to the X-ray LF of Hasinger et al. (2005; open points).  The 2SLAQ
  LF has been re-calculated in the redshift ranges presented by
  Hasinger et al.  The solid line denotes the LDDE model fit made by
  Hasinger et al. to their data.}
\label{fig:lfxray}
\end{figure}

We next compare to the X-ray luminosity function of
Hasinger et al. (2005), which is a soft X-ray LF of type 1 AGN from a
combined sample of $\sim1000$ objects in the $0.5-2.0$~\kev\ band.  We convert
from our $\mg(z=2)$ band to the $0.5-2.0$~\kev\ band, largely following
R05.  The $g$-band luminosity is given by
\begin{equation}
\log[L_{\rm g}(z=0)]=-0.4\mg(z=2)+20.638,
\end{equation}
assuming $\alpha_\nu=-0.3$.  We then convert to rest-frame luminosity
at 2500\AA\ via
\begin{equation}
\log(L_{2500})=\log[L_{\rm g}(z=0)]-0.3\log\left[\frac{4670}{2500}\right]
\end{equation}
and convert to mono-chromatic luminosity at 2\kev\ using
\begin{equation}
\log(L_{\rm 2\kev})=\log(L_{2500})+\alpha_{\rm
  ox}\log\left(\frac{\nu_{\rm 2\kev}}{\nu_{2500}}\right).
\end{equation}
The $\alpha_{\rm ox}$ parameter is the slope between 2500\AA\ and
2~\kev.  Various measurements show that this is dependent on
luminosity; we use the bisector method result of Steffen et
al. (2006), which gives
\begin{equation}
\alpha_{\rm ox}=-0.107\log(L_{2500})+1.740.
\end{equation}
Finally, we integrate over the $0.5-2.0$~\kev\ band assuming a photon
index of $\Gamma=2.0$.  This produces a final conversion of
\begin{equation}
\log(L_{\rm 0.5-2\kev})=36.972-0.288\mg.
\label{eq:lxmg}
\end{equation}
In Fig. \ref{fig:lfxray} we plot the comparison between the X-ray LF
of Hasinger et al. (2005) and our combined 2SLAQ+SDSS LF.  These LFs overlap in
the redshift intervals $0.4<z<0.8$ and $0.8<z<1.6$.  Overall, there is
impressive agreement between the optical and X-ray LFs.  This
agreement is best close to the break in the LF, with the X-ray LF
lying slightly above the optical LF at both fainter and brighter
magnitudes.  The bright end slope of the X-ray LF is somewhat flatter
than the optical LF.  We test how the luminosity dependence of
$\alpha_{\rm ox}$ impacts this by taking a generalize form of
Eq. \ref{eq:lxmg}: $\log(L_{\rm 0.5-2\kev})=A-B\mg$.  We fit for $A$
and $B$ by finding the values which bring the X-ray and optical LFs
into the closest agreement.  We find $B=0.362\pm0.023$, which infers a
significantly weaker luminosity dependence for $\alpha_{\rm ox}$ than
is found by Steffen et al. (2006).

\section{Model fits}\label{sec:mlfits}

\begin{table*}
\caption{The best fit pure luminosity evolution models
  (Eqs. \ref{eq:doublepl} and \ref{eq:mstarevol}). Listed are the
  redshift ranges and faint $\mg$ limit of the data fitted, the number of
QSOs in the analysis ($\nq$) and the best fit values of the model
parameters.  We also give the $\chi^2$ value for the comparison of the
model to the data, as well as the number of degrees of freedom ($\nu$)
and the $\chi^2$ probability.}
\label{tab:fitres}
\setlength{\tabcolsep}{2pt}
\begin{tabular}{@{}ccccccccccrrccc@{}}
\hline
Redshift& $\mg$  &  $\nq$ & $\alpha$ & $\beta$ & $\mg^*$ & $k_{1}$ & $k_{2}$ &  $\log(\Phi^{*})$ & $\chi^2$ & $\nu$ & $P_{\chi^2}$ & $D_{\rm KS}$ & $P_{\rm KS}$\\
range& limit & & & & & & &  ${\rm Mpc}^{-3}{\rm mag}^{-1}$ &\\
\hline 
0.4--2.1 & --21.5& 12977 & $-3.29\pm0.04$ & $-1.37\pm0.04$ & $-22.09\pm0.09$ & $1.44\pm0.03$ & $-0.319\pm0.011$ & $-5.79\pm0.07$ & 181.4 & 81 & 1.2e--09 & 0.014 & 6.2e--2\\ 
0.4--2.1 & --23.0& 11702 & $-3.38\pm0.05$ & $-1.48\pm0.04$ & $-22.36\pm0.10$ & $1.39\pm0.04$ & $-0.300\pm0.010$ & $-5.90\pm0.08$ & 152.1 & 72 & 1.1e--07 & 0.017 & 2.6e--2\\ 
0.4--2.3 & --21.5& 14146 & $-3.33\pm0.04$ & $-1.42\pm0.03$ & $-22.18\pm0.08$ & $1.44\pm0.03$ & $-0.315\pm0.008$ & $-5.84\pm0.07$ & 190.8 & 81 & 7.5e--11 & 0.019 & 1.4e--3\\ 
0.4--2.6 & --21.5& 15073 & $-3.33\pm0.04$ & $-1.41\pm0.03$ & $-22.17\pm0.08$ & $1.46\pm0.02$ & $-0.328\pm0.007$ & $-5.84\pm0.07$ & 256.6 & 80 & 2.2e--20 & 0.025 & 2.8e--5\\ 
\hline 
\end{tabular}
\end{table*}

As described above, the binned LF can be a biased estimator whenever the
LF changes significantly over the size of the bin (e.g. at the steep
bright end of the LF).  Therefore, a parametric QSO LF is also usually
derived by performing a maximum likelihood fit to the unbinned
data using the method first proposed by Marshall et al. (1983).  The
key issue in such fitting is to choose a functional form which is
representative of the data.  Optical QSO LFs have previously been well
fit by PLE models, and so we will start by considering this
parameterization.  Then we will investigate other forms such as
LDDE. 

The maximum likelihood approach has one disadvantage: the
normalization of the LF, $\Phi^*\equiv\Phi(\mg^*)$, cannot be
determined directly from the fitting.  We therefore estimate
this after the fitting process by integrating over the best fit model, such that
\begin{equation}
\Phi^*=\frac{N_{\rm Q}}{\int_{M_{\rm min}}^{M_{\rm max}}\int_{z_{\rm
      min}}^{z_{\rm max}}\frac{\Phi(M,z)}{\Phi^*}A(M)\frac{{\rm d} V}{{\rm d} z}{\rm d}M{\rm d}z}, 
\label{eq:phistar}
\end{equation}
where $N_{\rm Q}$ is the total number of QSOs, $A(M)$ is the effective area
of the survey (which is a function of magnitude if combining different
samples; i.e. 2SLAQ and SDSS), ${\rm d} V/{\rm d}
z$ is the cosmological volume element and $\Phi(M,z)/\Phi^*$
is the result of the ML fitting process.  The uncertainties in
$\Phi^*$ arise from the Poisson error in $N_{\rm Q}$ and the
uncertainties in the fitted model parameters; the latter is
the dominant term in our case.  The $\Phi^*$ error is given by the maximum
range of the $\Phi^*$ values from model parameter sets that are within
the $1\sigma$ N-dimensional confidence contours of the maximum
likelihood fit (e.g. for 5 parameters, within a region where
$\Delta\chi^2<5.89$).

A significant issue in the ML process is deciding on the
range of redshift and absolute magnitude over which to perform the fitting.  The
depth of the 2SLAQ data, combined with the detailed
completeness estimates at faint magnitudes, allows us to fit to fainter
limits than those used by C04 and R05 for the 2QZ and preliminary
2SLAQ analyses, respectively.  Therefore, we fit all the QSOs with  
$\mg(z=2)<-21.5$.  We will also test the impact of applying a limit of
$\mg(z=2)<-23.0$, which is approximately equivalent to the C04 and R05
limit.  In redshift we constrain our fitted range to be $z>0.4$, as at
lower redshift our photometric selection completeness is less certain
(see C09).  At the high redshift end we fit our data up to $z=2.6$; above
this redshift the photometric selection completeness drops below 50
per cent.  In order to fully constrain both the bright and faint end
of the QSO LF we fit the models to the combined 2SLAQ+SDSS LF.  

In assessing the goodness of fit for each model, we will measure the
$\chi^2$ value comparing the number of QSOs in a bin and the number
predicted by the best fit model (after accounting for
incompleteness).  As the unbinned maximum likelihood fitting method
makes use of a Poisson probability distribution function, the best fit
model will not necessarily correspond to that with the minimum
$\chi^2$ (which assumes a Gaussian probability distribution).  The
binning used to calculate our $\chi^2$ values can also have an impact
on the resulting values.  As we tend to smaller bins the estimated
numbers will be dominated by shot-noise.  With larger bins, containing
more QSOs, any systematic errors and/or mis-match of the model will
increase $\chi^2$.  As a result, the calculated $\chi^2$ values should
be treated as relative assessments of the goodness of fit, rather than
absolute ones.  We choose to use absolute magnitude bins with $\Delta
\mg=0.5$ and divide the LF into 6 uniform redshift intervals over the
range fitted.  As a second independent statistical test, which does
not depend on binning, we also apply a 2--D Kolmogorov--Smirnov (K--S)
test (Peacock 1983).  This test compares the model LF to the distribution
of QSOs in the $\mg-z$ plane.

\subsection{Pure luminosity evolution fits}\label{sec:plefits}

We assume the standard double power law of the form
\begin{equation}
\Phi(\mg,z)=\frac{\Phi(\mg^*)}{10^{0.4(\alpha+1)(\mg-\mg^*)}+10^{0.4(\beta+1)(\mg-\mg^*)}}, 
\label{eq:doublepl}
\end{equation}
where $\Phi$ is the comoving space density of QSOs.
The redshift dependence is characterized purely by evolution in
$\mg^*$.  We follow Boyle et al. (2000) by parameterizing this
evolution as a second order polynomial in redshift such that
\begin{equation}
\mg^*(z)=\mg^*(0)-2.5(k_1z+k_2z^2).
\label{eq:mstarevol}
\end{equation}
We note that this functional form for $\mg^*(z)$ requires symmetric
evolution about the brightest $\mg^*$ value.  This is likely to break down
at redshifts well above the peak (e.g. Richards et al. 2006), but our
sample is not able to probe to these redshifts.

The resulting best
fit parameters for the PLE models (Eqs. \ref{eq:doublepl} and
\ref{eq:mstarevol}) are listed in Table \ref{tab:fitres}.  The PLE
model is a relatively poor fit to the data at $\mg(z=2)<-21.5$ and $0.4<z<2.6$.
A $\chi^2$ comparison of the unbinned ML fit to the binned data (after
correcting for incompleteness in the bins) gives
$\chi^2/\nu=256.6.5/80$.  The most significant discrepancies between the
data and the model are at the faint end of the LF, particularly at
high redshift.  This is seen in Fig. \ref{fig:lfcomb} where the PLE
model (green dotted line) compared to the binned LF.  The qualitative
agreement actually appears good for a substantial fraction of the LF,
even though the overall agreement is poor due to the small statistical
errors on the LF measurements.
If we restrict the redshift range being
fit then we obtain a significant improvement, with $\chi^2/\nu$ of
$190.8/81$ for $0.4<z<2.3$ and $181.4/81$ for $0.4<z<2.1$.  Reducing
the magnitude range to $\mg<-23$ makes a further improvement to the
fitting, giving $\chi^2/\nu=152.1/72$.  The 2--D K--S tests show a
similar trend, although the K--S probabilities of acceptance are on
average higher.  The full redshift and magnitude range haves a K--S
probability of acceptance of only 2.8e--5, while the most restricted
data set has a K--S probability of $\simeq3$ per cent, and thus is
marginally acceptable (at the $\simeq2\sigma$ level).  One of the
reasons for
the poor fits is shown in Fig. \ref{fig:lfz}, where we find that the
space density of fainter QSOs peaks at lower redshift.

\subsection{Luminosity dependent density evolution fits}

We next investigate whether a luminosity dependent density evolution
(LDDE) model, as first suggested by Schmidt \& Green (1983), provides
an improvement in $\chi^2$ compared to PLE.  We use the model
described by Hasinger et al. (2005) and
others, which, when expressed in absolute magnitudes, takes the form
\begin{equation}
\Phi(\mg,z)=\frac{Ae_{\rm
  d}(\mg,z)}{10^{0.4(\alpha+1)(\mg-\mg^*)}+10^{0.4(\beta+1)(\mg-\mg^*)}},
\end{equation}  
where $A$ provides the normalization.  In this case $\mg^*$ does not
evolve, and the evolution is given by the term
\begin{equation}
e_{\rm d}(\mg,z)=\left\{ \begin{array}{ll}
(1+z)^{p_1}  & \mbox{if $z\leq \zc$,} \\ 
(1+\zc)^{p_1}[(1+z)/(1+\zc)]^{p_2} & \mbox{if $z> \zc$},
\end{array} \right.
\label{eq:ldde2}
\end{equation}
with
\begin{equation}
\zc(\mg)=\left\{ \begin{array}{ll}
\zcz 10^{-0.4\gamma(\mg-M_{\rm g,c})} & \mbox{if $\mg \geq M_{\rm g,c}$} ,\\
\zcz & \mbox{if $\mg < M_{\rm g,c}$}.
\end{array} \right.
\label{eq:ldde3}
\end{equation}
This model does not provide an improved fit (over PLE) to our combined
2SLAQ+SDSS data set.  Fitting over the range $0.4<z<2.6$ and
$\mg<-21.5$  gives a $\chi^2/\nu=256.6/77$ (8 free parameters), compared to
$\chi^2/\nu=256.6/80$ for a PLE model (5 free parameters) over the
same interval.  As suggested by  Hasinger et al. (2005), we also add a
luminosity dependent term to the power law exponents $p_1$ and 
$p_2$ in Eq \ref{eq:ldde2} such that
\begin{equation}
p_1(\mg)=p_{1,24}-\epsilon_1(\mg+24),
\label{eq:ldde4}
\end{equation}
\begin{equation}
p_2(\mg)=p_{2,24}-\epsilon_2(\mg+24),
\label{eq:ldde5}
\end{equation}
where the normalization is at $\mg=-24$.  A model fit which includes
these extra terms also fails to make a significant improvement on the
quality of fit ($\chi^2/\nu=255.6/75$).

When we compare the above LDDE model to the binned LF, a substantial
part of the disagreement appears to arise from the piecewise nature of
the functional form, which causes sudden changes in shape of the
model.  Therefore, we modify the above LDDE model so that the
functions describing the evolution are smoothly varying, rather than
the piecewise descriptions given by Eqs. \ref{eq:ldde2} and
\ref{eq:ldde3}.  The functional form we chose was similar in shape
to the piecewise model and described by 
\begin{equation}
e_{\rm d}(\mg,z)=\frac{2(1+\zc)^{p_1}}{[(1+z)/(1+\zc)]^{-p_1}+[(1+z)/(1+\zc)]^{-p_2}},
\label{eq:lddesmooth1}
\end{equation}
where
\begin{equation}
\zc(\mg)=\frac{\zcz}{1+10^{0.4\gamma(\mg-M_{\rm g,c})}}.
\label{eq:lddesmooth2}
\end{equation}
This form of the LDDE model provides a much improved fit over the
piecewise form used by H05 (with the same number of degrees of
freedom).  We find $\chi^2/\nu=146.8.3/77$ when fitting to QSOs at
$0.4<z<2.6$ and $\mg<-21.5$.  The K--S test suggests that the model
and data are in agreement at the $\sim2$ per cent level, although the
$\chi^2$ test rejects the model at much higher significance.
As above, including further
terms to allow $p_1$ and $p_2$ to vary with magnitude
(Eqs. \ref{eq:ldde4} and \ref{eq:ldde5}) does not improve $\chi^2$.
The best fit smooth LDDE model parameters are listed in Table
\ref{tab:ldde} and the model is plotted in Figs. \ref{fig:lfcomb} and
\ref{fig:lfz} (long dashed lines).  From Fig. \ref{fig:lfz} we can see
that the advantage this model has over PLE is that the space density
of fainter sources peaks at lower redshift.  In this model the faint end slope 
is not fixed, but varies as a function of both redshift and
luminosity.  This causes the dip and then upturn seen at $z>1$ in
Fig. \ref{fig:lfcomb} beyond the range of the data.  Such changes in
shape beyond the limits of the data should not be taken seriously and
are unlikely to be real.

\begin{table}
\caption{The best fit LDDE model for $0.4<z<2.6$ and $\mg(z=2)<-21.5$.
  15073 QSOs were used in the fit.  
We also give the results of comparing the model to the data via
  $\chi^2$ and K--S tests, and include the number of degrees of
  freedom ($\nu$).}
\label{tab:ldde}
\centering
\begin{tabular}{cc}
\hline
Parameter & Value\\
\hline 
$\alpha$ & $-3.70\pm0.06$\\
$\beta$ &  $-2.34\pm0.03$ \\
$\mg^*$ & $-26.69\pm0.16$\\
$M_{\rm g,c}$ &         $-23.90\pm0.14$\\ 	
$\gamma$ &	   $0.68\pm0.02$\\
$z_{\rm c,0}$ & 	   $2.47\pm0.05$\\	
$p_1$ & 	   $6.28\pm0.18$\\ 
$p_2$ & 	   $-2.85\pm0.21$\\ 
$\log(A)$&  $-9.21\pm0.18$\\ 	
 $\chi^2$ & 	   146.8\\ 		
 $\nu$ &	   77\\ 		
 $P_{\chi^2}$ &	   2.9e--6 \\       
 $D_{\rm KS}$ & 0.015\\
 $P_{\rm KS}$ & 1.7e--2\\

\hline
\end{tabular}
\end{table} 

\subsection{Fitting narrow $\mg$ and $z$ intervals}

\begin{figure}
\centering
\centerline{\psfig{file=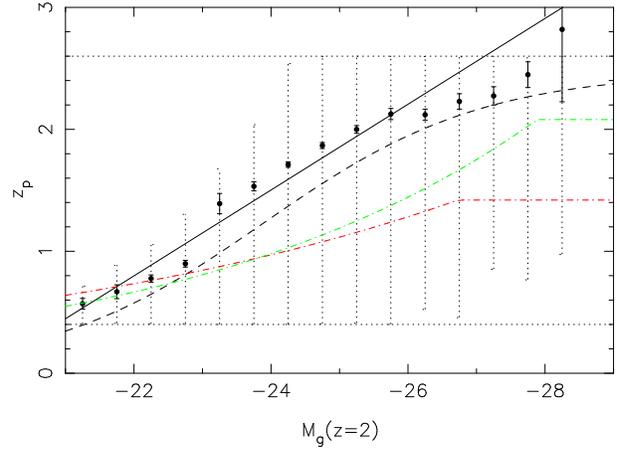,width=8cm,angle=270}}
\caption{The redshift, $z_{\rm p}$, at which the space density of QSOs peaks as a
  function of $\mg$.  This is determined by fitting Eq. \ref{eq:singlpl} to our data in
  narrow slices of full width $\Delta\mg=0.5$ mags.  The horizontal
  dotted lines indicate the 
  nominal redshift range that was fitted over ($z=0.4-2.6$), while the
  vertical dotted lines show the actual redshift range for each $\mg$
  bin.  The solid line shows the best linear fit and the dashed
  line is the best fit $z_{\rm c}(\mg)$ using the LDDE model
  (Eq. \ref{eq:lddesmooth2}).  The dot--dashed lines show the
  evolution of $z_{\rm c}$ found by Hasinger et al. (2005; red
  dot--dashed line) and Bongiorno et al. (2007; green dot--dashed
  line) when fitting a piecewise form of the LDDE model to X-ray and
  optical data respectively.}
\label{fig:zpeak}
\end{figure}

\begin{figure*}
\centering
\centerline{\psfig{file=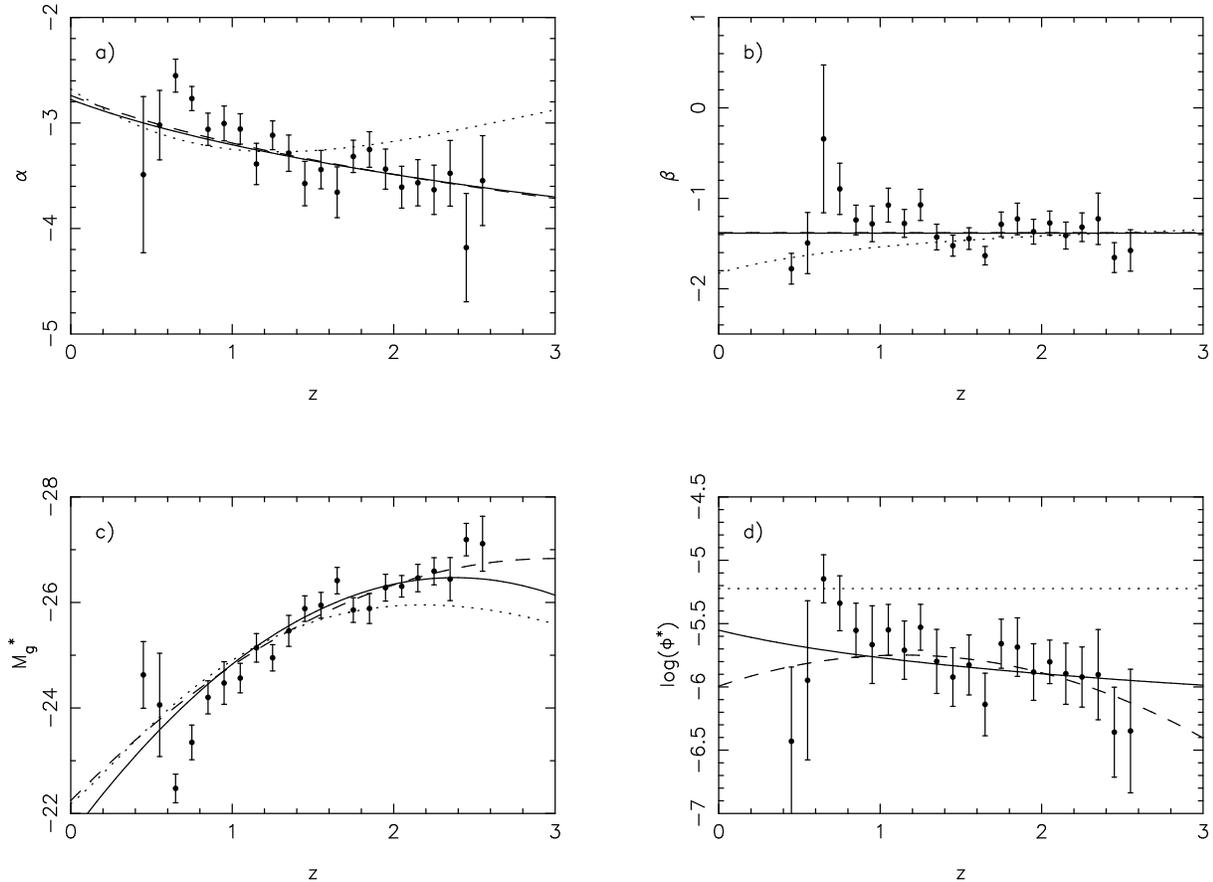,width=16cm,angle=270}}
\caption{Luminosity function parameters for fits of a double power law
  in narrow redshift slices. a) The bright end slope, $\alpha$, which
  shows a significant steepening towards high redshift.  At redshifts
  higher than $z\sim3$ the bright end slope has been shown by R06 to
  flatten again. b) The faint
  end slope, $\beta$, which shows no significant change as a function of redshift. c)
  $\mg^*$, which shows the well-known strong brightening towards high
  redshift.  d) the normalization, $\Phi^*$, which shows a systematic reduction
  towards high redshift.  The dotted lines show the best fit models derived by Hopkins
  et al. (2007).  The solid and dashed lines show the best fits from
  our LEDE model, assuming power law and quadratic evolution for
  $\Phi^*$ respectively.}
\label{fig:lfparams}
\end{figure*}

While the LDDE model provides an improved fit over PLE,
we now explore different forms of the QSO LF which may provide a
better match to our data.
We start by using the unbinned ML
approach to fit the space density of QSOs as a function of redshift in
narrow $\mg$ slices of full width 
$\Delta\mg=0.5$.  A single power law in luminosity with quadratic density evolution
is used,
\begin{equation}
\Phi(\mg,z)=\frac{\Phi^*10^{(Az(1-0.5z/z_{\rm p}))}}{10^{0.4(\alpha+1)\mg}},
\label{eq:singlpl}
\end{equation}
where we fit for $\alpha$, $A$ and $z_{\rm p}$.  For a single power
law the quadratic
evolution in density is equivalent to the quadratic evolution in
$\mg^*$ described by Eq. \ref{eq:mstarevol}.  We have re-parameterized
this so that one of the parameters ($z_{\rm p}$) gives the redshift of
the peak of the QSO space density.  This allows measurement of
 the peak redshift, $z_{\rm p}$, as a function of $\mg$ which is 
  plotted in Fig \ref{fig:zpeak}.  This figure shows quantitatively the trend
  that is apparent in  Fig. \ref{fig:lfz} and demonstrates that we see
  AGN downsizing at high significance.  At the extremes of
 the $\mg$ distribution the restricted redshift range
 of the data (due to the flux limited nature of the sample, see dotted
 lines in Fig. \ref{fig:zpeak}) may be
 biasing our estimate of the peak redshift, but brighter than
 $\mg=-24$, where the redshift coverage is uniform, we see a 
 consistent trend of increasing $z_{\rm p}$ with luminosity.
  The best fit linear
 relation is $z_{\rm p}=(-6.94\pm0.16)+(-0.352\pm0.007)\mg$, so the
 gradient is non-zero at a highly significant level.  A Spearman rank
 correlation test finds that $z_{\rm p}$ and $\mg$ are
 correlated at 99.98 per cent significance.  The low luminosity AGN
  with $\mg>-23$ peak in space density below $z\simeq1$ and the
  peak redshift increases monotonically up to the most luminous
  sources we sample.  At $\mg>-26$
 the points are consistently below the best fit linear relation.  This
 corresponds to $z\simeq2.2$, but may be biased by the
 $z\simeq2.6$ redshift cut off in our sample.  The reality of such a
 flattening of the peak redshift could be confirmed by faint QSO
 samples that probe to higher redshift, such as the
 AAOmega-UKIDSS-SDSS (AUS) Survey currently underway at the
 Anglo-Australian Telescope (Croom et al. in preparation).  In
 Fig. \ref{fig:zpeak} we also plot $z_{\rm c}(\mg)$ from the LDDE
 model (Eq. \ref{eq:lddesmooth2}).  This shows increasing peak
 redshift for brighter QSOs, although $z_{\rm c}(\mg)$
 is typically between 0.2 and 0.4 lower than the directly measured
 $z_p$.  This is due to the different functional forms fit in each
 case (quadratic vs. double power law).

If the completeness of the 2SLAQ sample was overestimated at the
faintest magnitudes, this could mimic
downsizing.  In order to examine this possibility we re-determine the
binned LF, but applying a faint flux limit of $g=21.0$ to our sample.
The LF limited at brighter magnitudes is
completely consistent with that derived from the full sample.  Indeed,
the magnitude intervals at $\mg=-24.5$ and $-25.5$ still show a
distinct flattening and turn-over (Fig. \ref{fig:lfz}), which is not seen at brighter
absolute magnitudes.

We next fit a simple double 
power law (i.e. Eq. \ref{eq:doublepl}) to the data in narrow redshift
intervals, $\Delta z=0.1$.  Over this narrow range we do not allow for
any evolution, and so fit for only three parameters: $\alpha$, $\beta$ and
$M^*$.  We also derive a fourth parameter, $\Phi^*$, which cannot be
fit for using the ML method, but is derived from Eq. \ref{eq:phistar}.
The fitted parameters are shown as a 
function of redshift in Fig. \ref{fig:lfparams}.  There is covariance
between the fitted parameters for a given redshift 
slice. However, for a given parameter the measurements between
different redshift slices are independent.
Fig. \ref{fig:lfparams} shows significant trends with redshift which
are inconsistent with PLE.  In Fig. \ref{fig:lfparams}a the
bright end slope, $\alpha$, shows significant (99.9 per cent from a Spearman rank
test) steepening with increased redshift,
in agreement with previous results (e.g. Goldschmidt \& Miller 1998;
C04; R06). Hopkins et al. (2007) also find a significant change of
$\alpha$ at $z~\simlt~2$, which is described by the dotted line in
Fig. \ref{fig:lfparams}a.  This is inconsistent with our measurement
of the evolution of $\alpha$ at the 98 per cent level (via a $\chi^2$
test between the Hopkins et al. model and our data points), although
it shows the same general trend.  The
faint end slope ($\beta$; Fig. \ref{fig:lfparams}b) is consistent with
no evolution; this is in contrast to the measurement of Hopkins et
al. (2007).  This difference may be due to our redshift limit of
$z>0.4$, as Hopkins et al. find the steepest values of $\beta$ below
this redshift.  Again, a $\chi^2$ test between our data and the
Hopkins et al. model is inconsistent at the 98 per cent level.  The
best fit value of $\mg^*$ (Fig. \ref{fig:lfparams}c) shows
the strong evolution expected in a PLE model, but we also see a
systematic decline in $\Phi^*$ (Fig. \ref{fig:lfparams}d) which is not
part of the standard PLE model.  The correlation between $\Phi^*$ and
$z$ is significant at the 94 per cent level from a Spearman-rank
correlation test.  The lowest two redshift bins in
Fig. \ref{fig:lfparams} appear to have fitted values which lie off the
trend defined by the values at other 
redshifts; $\Phi^*$ is lower,  $M^*$ is brighter and $\alpha$
is steeper.  If we ignore these two lowest redshift bins, then the
significance of the correlation between $\Phi^*$ and $z$ increases to
99.96 per cent.  The above results suggest several modifications to the
PLE model which should improve the fit of the model; we investigate
this possibility below.  The Hopkins et al.\ model for $\mg^*$ and
$\Phi^*$ is also plotted (dotted lines) in Figs. \ref{fig:lfparams}c
and d.  While the overall trend for $\mg^*$ is the same, the Hopkins
et al. model is somewhat flatter.  The Hopkins et al.\ fit for $\Phi^*$
is systematically $\simeq0.4$ dex higher than our measurements (after
converting from ${\rm d}\Phi/{\rm d}\log(L)$ to ${\rm d}\Phi/{\rm
  d}M$).  A possible cause of this offset is that Hopkins et al.\
fit their models to a broad range of {\it binned} data, rather than
the unbinned model fits carried out here.

\subsection{Modified PLE fits}\label{sec:ledefits} 

 The
 first modification we make to the standard PLE model described in
 Section \ref{sec:plefits} is to allow the bright end slope, $\alpha$, to vary
 with redshift.  We follow a parameterization similar to that of Hopkins et al.
 (2007) and use
\begin{equation}
\alpha(z) = \alpha_{\rm ref}\left(\frac{1+z}{1+z_{\rm ref}}\right)^{p_\alpha},
\label{eq:alphaevol}
\end{equation}
where $z_{\rm ref}$ is fixed at $z_{\rm ref}=2$.  Hopkins et al.\ (2007)
used a double power law to track the flattening of $\alpha$ at $z>3$.
As our sample does not probe to this redshift, we use a single
power law parameterization.

We account for the evolution in $\Phi^*$ seen in
Fig. \ref{fig:lfparams}d using a power law parameterization such that
\begin{equation}
\Phi^*(z) = \Phi^*_{\rm ref}\left(\frac{1+z}{1+z_{\rm ref}}\right)^{p_\Phi}.
\end{equation}   
Again, we take $z_{\rm ref}=2$.
Fitting this model over the full redshift range ($0.4<z<2.6$) and at
$\mg<-21.5$ results in a $\chi^2/\nu=213.3/78$ and $P_{\rm
  KS}$=1.9e--4.  This is an improvement over PLE, but not as good as
the LDDE model.  In
Fig. \ref{fig:lfparams} we compare this model
fit (solid lines) to the dependencies of $\alpha$, $\beta$, $\mg*$ and
$\Phi^*$ with redshift.  This model provides a much improved description of the evolving
bright end slope, $\alpha$ and normalization, $\Phi^*$.  In order to
further improve the model fit, we try a different parameterization for
the evolution in $\Phi^*$.  This is a quadratic form, similar to the
evolution of $M^*$ in Eq. \ref{eq:mstarevol}, such that 
\begin{equation}
\log(\Phi^*) = \log(\Phi^*_0)+(k_{\Phi1}z(1.0-0.5z/k_{\Phi2})).
\label{eq:phievol}
\end{equation}
Adding this functional form allows a significant improvement in the
fit over the full redshift range, with $\chi^2/\nu=121.0/77$ and
$P_{\rm KS}=0.30$.  We call
this final model luminosity evolution + density evolution (LEDE) and it is
plotted in Figs. \ref{fig:lfcomb} and \ref{fig:lfz} (short dashed
lines).  The most noticeable differences between this model and PLE is
the change in amplitude at high redshift and the bright end slope
change.  The evolution of $\Phi^*$ has the effect of shifting the peak
space density of low luminosity QSOs towards lower redshift,
i.e. downsizing (see Fig. \ref{fig:lfz}).  The results of the LEDE fit
are also shown as the dashed lines in
Fig. \ref{fig:lfparams}.  The evolution in $\Phi^*$ is best fit by a
quadratic which is convex; that is, it declines at both low and high
redshift.  The power law and quadratic forms for the
evolution in $\Phi^*$ are in reasonable agreement over most of the
redshift range fitted.  However, outside of this range they diverge
markedly.  This highlights the danger of extrapolating such
empirically derived functional forms outside of the redshift and
luminosity ranges over which they are fitted.  In particular, it is well
known that at $z>3$ the bright end of the QSO LF flattens again, which
is not accounted for in our model.

\begin{table}
\caption{The best fit modified LEDE model (i.e. using
  Eqs. \ref{eq:alphaevol} and \ref{eq:phievol}) for $0.4<z<2.6$ and
  $\mg(z=2)<-21.5$.  15073 QSOs were used in
  the fit.  We also give the results of comparing the model to the data via
  $\chi^2$ and K--S tests, and include the number of degrees of
  freedom ($\nu$).} 
\label{tab:fitlede}
\centering
\begin{tabular}{cc}
\hline
Parameter & Value\\
\hline 
$\alpha_{\rm ref}$ & $-3.48\pm0.05$\\
 $p_\alpha$ & 	   $0.220\pm0.018$\\	
$\beta$ &  $-1.38\pm0.03$ \\
$\mg^*$ & $-22.24\pm0.09$\\
 $k_{1}$ &         $1.23\pm0.03$\\ 	
 $k_{2}$ &	   $-0.206\pm0.007$\\
$k_{\Phi1}$ & 	   $0.430\pm0.034$\\	
$k_{\Phi2}$ & 	   $1.139\pm0.034$\\ 
$\log(\Phi^{*}_0)$&  $-5.79\pm0.07$\\ 	
 $\chi^2$ & 	   121.0\\ 		
 $\nu$ &	   77\\ 		
 $P_{\chi^2}$ &	   1.0e--3 \\       
 $D_{\rm KS}$ & 0.0096 \\
 $P_{\rm KS}$ & 0.30 \\
\hline 
\end{tabular}
\end{table} 

\section{Discussion}\label{sec:dis}

\subsection{Comparison of number counts to 2QZ}\label{sec:disnm}

The number counts and luminosity function from 2SLAQ broadly agree
with other works, but provide a significant advance in the precision
available to constrain the faint end of the QSO LF at $z<2.6$.  We
confirm that the 2SLAQ survey sees an excess in counts over the 2QZ
fainter than $g\simeq20.0$.  To further examine this, we re-calculate
the 2SLAQ number counts 
in identical bins to 2QZ (correcting the difference in pass-bands
using $g-\bj=-0.045$).  Brighter than $g\simeq20.0$ the 2QZ
and 2SLAQ QSO number counts agree well.  However, we find that the
ratios of the differential counts are $N_{\rm 2QZ}(g)/N_{\rm
  2SLAQ}(g)=0.85\pm0.03$ and $0.75\pm0.02$ for 0.25 mag bins centred
on $g=20.38$ and $g=20.63$ respectively.  The
slope of the integrated 2SLAQ number counts at this magnitude is
$\simeq0.36$.  Thus, a systematic offset in magnitude of $\Delta
g\simeq0.35$ would be sufficient to cause this difference.  Such an
error must be magnitude dependent, as there is no visible offset between
2QZ and 2SLAQ brighter than $g\simeq20$.

\begin{figure}
\centering
\centerline{\psfig{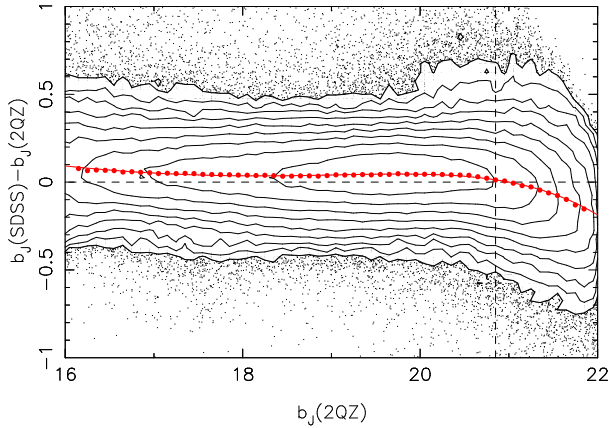}}
\caption{The difference in stellar magnitudes between SDSS and 2QZ in
  the $\bj$ band, as a function of $\bj$.  The small points and
  contours show the distribution of $\simeq600,000$ stars in the 2QZ
  NGP region which also have SDSS photometry.  The contours are
  spaced logarithmically by 0.25 dex.  The vertical dashed line marks
  the faint limit of the 2QZ survey at $\bj=20.85$.  The large red
  points show the median $\bj(SDSS)-\bj(2QZ)$ at 0.1 mag intervals.
  The solid red line shows the best fit polynomial to these points.}
\label{fig:2qzsdssphot}
\end{figure}

There are several possible causes for the observed deficit of
faint 2QZ QSOs.  First we check for scale errors in the calibrated
photographic photometry, by cross matching the 2QZ input catalogue of
stellar objects (Smith et al.\ 2005) with SDSS imaging data, finding
$\sim600,000$ matches to $\bj\simeq22$.  This is only
possible in the NGP region of the 2QZ (which also overlaps with
2SLAQ).  The SDSS magnitudes are converted to an effective SDSS $\bj$
magnitude by combining the relation $\bj=B-0.28(B-V)$ 
\cite{bg82} with the SDSS colour transformations given by Jester et
al. (2005) to give
\begin{equation}
\bj=g+0.116(g-r)+0.148.
\label{eq:jesterstars}
\end{equation}
This transformation is appropriate for stars, as we are comparing the
photometry from the full stellar 2QZ catalogue, not just the objects
selected as QSO candidates.  A slightly different transformation is
appropriate for QSOs (see Jester et al. 2005).  We calculate the median
magnitude difference
[$\bj(SDSS)-\bj(2QZ)$] in 0.1 magnitude intervals, as is shown
in Fig. \ref{fig:2qzsdssphot}.  This difference between the 2QZ and
SDSS magnitudes is almost constant from $\bj=16$ to 20.85 and is well
described by a 8th order polynomial (solid red line) over this range.
At $\bj=18-20$, $\bj(SDSS)-\bj(2QZ)$ is roughly constant, $\simeq0.04$ mags, and
declines to $\simeq0.01$ mag at $\bj=20.85$.  This
$\simeq0.03$ scale error between $\bj=20$ and $20.85$ is in the right
sense to explain the number counts discrepancy, but is an order of
magnitude too small.  

A second possible cause of the number counts difference is that the
colour selection 
of the 2QZ is less complete than was estimated by C04.  We match the
2SLAQ QSOs to the 2QZ photometry and find that 95 per cent of 2SLAQ
QSOs (at $0.4<z<2.1$) would have been selected by 2QZ, independent of
$\bj$. 
This rules out colour selection as the cause of the number
counts discrepancy.  It could also be the case that as we approach the
plate limit, sources are not being detected on the UKST plates.
When matching 2SLAQ QSOs to the 2QZ photometry, we find that brighter than
$g=20$, $93\pm3$ per cent of 2SLAQ sources can be matched to the 2QZ
stellar photometry (those missing are in large part due to holes
around bright stars in the 2QZ catalogue).  At $g=20.4$ to $20.8$ only
$88\pm3$ per cent are matched to 2QZ, indicating that an increasing
fraction of sources are missing from the 2QZ catalogue at fainter
fluxes, although again, the effect seen is not sufficient to explain
the observed discrepancy in the number counts.  We suspect that a
combination of small photometric calibration errors, missing objects
at fainter fluxes and other currently unknown errors together
contribute to the discrepancy seen between the 2SLAQ and 2QZ number
counts.

\subsection{LF models, evolution and downsizing}

The break in the QSO luminosity function we've measured in this paper is a
gradual flattening which takes place over several magnitudes.  Most of
this flattening occurs just faint-wards of the SDSS QSO LF, but in the
region of overlap, the SDSS and 2SLAQ LFs are in excellent agreement.
The combination of SDSS and 2SLAQ allows us to simultaneously place
accurate constraints on both the bright and faint ends of the QSO LF,
with errors in the binned LF typically $<0.1$ dex (and often $\simeq0.02-0.03$ dex)
over a range of 5--6 magnitudes (i.e. over a factor of 100 in
luminosity).  With errors of this size much greater care needs to be
taken over systematic errors, as these may dominate over the
statistical errors.  We find that
previous smaller surveys generally agree with the higher precision
2SLAQ+SDSS LF measurements. 

\begin{figure}
\centering
\centerline{\psfig{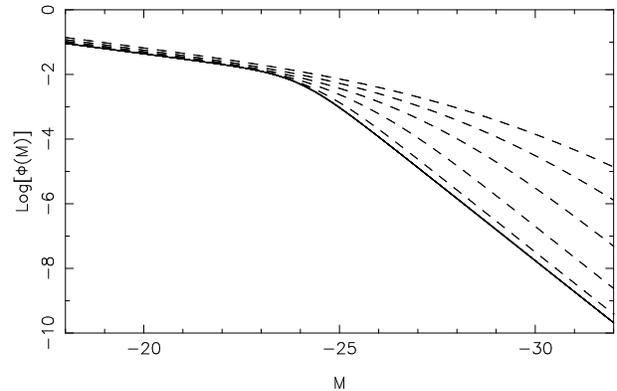}}
\caption{The effect of convolving a double power law luminosity
  function with a Gaussian distribution in magnitude.  The original
  double power law (solid line) is convolved with Gaussians of five
  different widths, $\sigma=0.5$, 1.0, 1.5, 2.0, and 2.5 mags (dashed lines, bottom
  to top).}
\label{fig:lfconv}
\end{figure}

When comparing our combined QSO LF to the soft X-ray LF of type 1 AGN
measured by Hasinger et al.\ (2005) we find good agreement near
$M^*$.  However, the measured bright end slope is significantly  
steeper in the 2SLAQ+SDSS data set than the X-ray data.  We find that
an $\alpha_{\rm ox}$ which has a much weaker dependence on luminosity
than previously measured (e.g. Steffen et al. 2006) provides
an improved match between the luminosity functions.  Green et
al. (2009) find a weaker luminosity dependence for $\alpha_{\rm ox}$,
but this is largely limited to fainter luminosities ($M_{\rm B}>-23$),
and they find a result similar to Steffen et al. at brighter
magnitudes.  The measured RMS scatter in $\alpha_{\rm ox}$ could
potentially alter the shape of the LF.  In Fig. \ref{fig:lfconv} we
plot a fiducial double power law luminosity function (solid line) and
the convolution of this with a Gaussian
distribution (dashed lines).  The LF retains its shape well away from
the break, but the bright end slope is flattened near the break.
The scatter in $\log(L_{\rm x})$ found by Steffen et al. (2006) is
$\sim0.3-0.4$ dex, corresponding to $\sim0.75-1.0$ mags.  Such a
scatter provides qualitatively the correct flattening of the bright
end of the QSO LF, although only locally, within $\sim2-3$ mags of
$M^*$.  Of course, the probability distribution function for such a
scatter may not be Gaussian, and may also depend on luminosity.
Detailed consideration of this is outside the scope of the current
paper, and would ideally require the measurement of the bivariate
X-ray/optical LF for a sample that was largely complete at both
wavelengths.  However, we do note that if scatter in $\alpha_{\rm ox}$
is the cause of the LF difference, this infers that optical/UV
luminosity is the independent variable, and X-ray luminosity is the
dependent variable.  That is, we need to take the optical LF and
convolve it with a Gaussian to match the two, rather than take the
X-ray LF and convolve this with a Gaussian. 
This is as one might expect, given that the
optical/UV light originates from the accretion disk, while the X-ray
originates from a hot corona above the disk.  In fact, it could be
argued that both the optical/UV and X-ray luminosities are dependent on
the bolometric luminosity, which should directly correlate to the
accretion rate.  In this case, the dispersion in the  bolometric
correction to the X-ray must be larger than the dispersion in the
bolometric correction to the UV/optical.  

The results of Mahony et
al. (2009) suggest a second possible explanation of the
flatter bright end slope of the X-ray LF.  They find that the fraction
of AGN in the {\it ROSAT} All Sky Survey Bright Source Catalogue (BSC)
which are detected at radio wavelengths by the NRAO VLA Sky Survey
\cite{nvss} and the
Sydney University Molonglo Sky Survey \cite{sumss} increases with increasing
redshift.  At $z>1$ almost all QSOs in the BSC are detected in these
radio surveys.  The
natural explanation of this is that the very luminous high redshift
QSOs in the sample are having their X-ray flux boosted by a jet
component, which may also be Doppler boosted by beaming.  In contrast
the radio detected fraction (to the same radio flux limit) of optically
bright QSOs is no more that 25 per cent \cite{jiang07}.  Therefore,
the numbers of QSOs at the bright end of the X-ray LF will be boosted
relative to the number in the optical.

The qualitative impression of the LF evolution (seen in
Figs. \ref{fig:lfsdss} and \ref{fig:lfcomb}) is of a consistent shape
where the characteristic luminosity shifts with redshift.  This is the
classical pure luminosity evolution (PLE) seen in previous optical samples.
A very different impression is obtained from Fig. \ref{fig:lfz}.  Here
we see that the space density of fainter QSOs peaks at lower redshift
than that of high luminosity QSOs.  This is equivalent to the AGN
downsizing seen in X-ray samples (e.g. Barger et al. 2005), although
it has not been convincingly seen in optical samples until now.
Bongiorno et al. (2007) find that their optical LF is better fit by a
LDDE model than PLE, but do not provide any errors on their model fits
to directly assess the significance of downsizing.

Such downsizing immediately rules out PLE, which due to its functional
form, peaks in space density at the same redshift for every
luminosity.  Direct fitting of models to the unbinned QSO LF confirms
the disagreement with PLE, although we note that the best fit PLE
model plotted in Fig. \ref{fig:lfcomb} is relatively close to the data
in all but the highest redshift interval.  The disagreement with PLE is
much clearer in Fig. \ref{fig:lfz}.  The discord between PLE and the
2SLAQ LF is most prominent at low luminosities and high redshifts.
This is exactly the point at which the downsizing is most noticeable in
the binned LF.  There is also some disagreement at $z\sim1.5$ and
$\mg\sim-24$, where the errors on the LF are particularly small given
the large numbers of QSOs per bin.

A much improved fit is obtained if we use a LDDE model.  While the
functional form of Hasinger et al. (2005) does not provide a reduction
in $\chi^2$, if we modify this to use a smooth functional form (Eqs. 
\ref{eq:lddesmooth1} and \ref{eq:lddesmooth2}) a much better fit is
obtained.  However, this is still a relatively poor fit
($\chi^2/\nu=146.8/77$) and also behaves badly outside of the fitted
range of the data.  A model which combines luminosity evolution and
simple density evolution (our LEDE model) provides a further
improvement to the fit with $\chi^2/\nu=121.0/77$ ($P_{\chi^2}=0.001$)
and a K--S test which is formally consistent at the 1$\sigma$ level
($P_{\rm KS}=0.30$).  The LEDE model is the best match to our combined
2SLAQ+SDSS data set for both the $\chi^2$ and K--S tests.  Even in
this case the $\chi^2$ is still formally a poor fit, 
it is worth considering whether this is due to
residual systematic errors.  To do this we carry out the binned
$\chi^2$ test on the best fit models, adding a fixed fractional error
in quadrature to the statistical error on $\Phi$ in each bin.  We then
determine the systematic error 
required to make the models acceptable at the 5 per cent level
(i.e. $P_{\chi^2}=0.05$).  For the PLE model, a global 12 per cent
systematic error is required to achieve an acceptable match, while for
the LDDE and LEDE models this is only 6 and 4 per cent respectively.

If we fit the LF in narrow magnitude intervals we find further
evidence for downsizing (e.g. Fig. \ref{fig:zpeak}).  Even at
relatively bright magnitudes ($\mg>-24$) the redshift at which the
space density of QSOs peaks increases with luminosity, and the peak is
well below the maximum redshift of our sample, where completeness
starts to decline.  We find this trend is similar to that found by the
fit of the LDDE model to the full sample (dashed black line in
Fig. \ref{fig:zpeak}).  In comparison, the trends found by Bongiorno
et al. (2007; Fig. \ref{fig:zpeak}, green dot--dashed line) and
Hasinger et al. (2005; Fig. \ref{fig:zpeak}, red dot--dashed line),
fitting a similar LDDE model to VVDS and X-ray data respectively, were
somewhat flatter, although it is not clear whether this difference is
significant.  The maximum
$z_{\rm c}$ in the best fit model of Hasinger et al. (2005) is only
$z=1.42\pm0.11$ which disagrees with the peak in the space density of
high luminosity optical QSOs (e.g. R06 and this work).  Such a
disagreement is likely to be due to the relatively low number of X-ray
objects at the brightest luminosities.  At the lowest luminosities
there is good agreement between our estimated $z_{\rm p}$ and those
estimated from the other works.  There are various other evolutionary 
trends which are inconsistent with the simple PLE model; these include
evolution in the bright end slope $\alpha$ and evolution in $\Phi^*$
(see Fig. \ref{fig:lfparams}).

Recent simulations (e.g. Hopkins et al. 2006) suggest that the faint
end slope of the QSO LF is set by the light curves of QSOs fainter
than their peak luminosity.  These hydrodynamical simulations find
that the faint end slope is a function of peak luminosity and,
indirectly, a function of redshift, as the distribution of peak 
luminosities shifts towards higher luminosity at higher redshift (at
least up to $z\simeq2.5$).  Our best fit models show no evidence of
faint end slope evolution (e.g. Fig. \ref{fig:lfparams}) and find a
slope of $\beta=-1.38\pm0.03$.  However, this is consistent with the
model faint end slopes of Hopkins et al. at redshift $\simgt0.5$, as it
is only at the 
lowest redshifts that the faint end slope is predicted to evolve
appreciably.  In fact, the lowest redshift bins in
Fig. \ref{fig:lfparams}b do show a steeper slope, more consistent with
the predicted turn up (see Fig. 3 of Hopkins et al. 2006).

Further developments by Hopkins et al. (2008) placed the above light
curves into a cosmological context, combining evolving dark matter
halo mass functions, halo occupation distributions and merger rates to
predict the evolution of the QSO population.  This model gives
reasonably good agreement with a combination of recent LF
measurements, improving on earlier models (e.g. Wyithe \&
Loeb 2002).  The LF presented here has the best
combination of dynamic range and precision of any yet measured.  As
such, it will provide further constraints on QSO formation models.

We find highly significant (99.9 per cent) evolution of the bright end
slope of the LF, which steepens from $\alpha\simeq-3.0$ at $z\sim0$ to
$\alpha\simeq-3.5$ at $z\sim2.5$.  This strengthens the previous
evidence for such a trend (e.g. Goldschmidt \& Miller 1998), and is in
general agreement with the evolution seen by Hopkins et al. (2007),
although in detail the form of the evolution is somewhat different
(see Fig. \ref{fig:lfparams}a).  A naive direct mapping from the evolving
dark matter halo mass function would also produce such an evolution in
the bright end slope.  QSOs typically populate similar mass dark
matter halos at all redshifts $z<2.5$ (e.g. Croom et al. 2005).
Therefore, as the mass function becomes steeper at higher mass
(relative to the break in the mass function) and the break in the mass
function moves to lower mass at higher redshift, the typical QSO host
mass moves to a steeper part of the mass function with increasing
redshift.  However, at
$z>2.5$ the bright end of the LF is seen to flatten again (Richards et
al. 2006), in disagreement with a naive mapping from the halo mass
function.  The more complex models of Hopkins et al. (2008) appear to
reproduce such trends.

\section{Summary}\label{sec:sum}

In this paper we present the optical QSO LF with unprecedented
precision and
dynamic range.  We do this by combining the 2SLAQ and SDSS data sets
to probe both the faint and bright ends of the LF at $z<2.6$.
Although the evolution of QSO LF appears very similar to PLE, we find
significant departures from this form of evolution.  A form of LDDE
provides a better fit to the LF, but we find that this can still be
improved upon.  We find that the bright end slope and $\Phi^*$  both
show significant evolution, so use a modified PLE model with added
density evolution which we call luminosity evolution + density
evolution (LEDE).  The LEDE model produces the
best fit of all models investigated, although a systematic error of 4 per cent is
required to make the data formally consistent (at the $2\sigma$ level)
with the model in our $\chi^2$ test.  To make further progress in our
understanding of the QSO LF, new measurements at faint
magnitudes and high redshift ($z>3$) need to be made, making an
accurate measurement of the faint end slope and better constraining
the bright end slope.  The AAOmega-UKIDSS-SDSS (AUS) survey (Croom et
al., in preparation), currently underway on the Anglo-Australian
Telescope, aims to do this, reaching an equivalent magnitude limit to
2SLAQ, but up to $z\sim5.5$.   As improvements are made to photometric QSO
samples, they will also provide the opportunity to investigate the
faint end of the QSO LF (e.g. Richards et al. 2009).  Further progress
in the optical is also dependent on our ability to account for the
contamination in optical samples, particularly at redder wavelengths.
If we can do this, then new larger area optical imaging surveys
(e.g. LSST; Ivezi\'{c} et al. 2008) will allow substantial improvements in
our characterizing of the evolution of AGN.

\section*{ACKNOWLEDGEMENTS} 

The 2SLAQ Survey is based on observations made with the
Anglo-Australian Telescope and the Sloan Digital Sky Survey.
We warmly thank all the present and former staff of the
Anglo-Australian Observatory for their work in building and operating
the 2dF facility.  The 2QZ is based on observations
made with the Anglo-Australian Telescope and the UK Schmidt Telescope.

Funding for the SDSS and SDSS-II has been provided by the Alfred
P. Sloan Foundation, the Participating Institutions, the National
Science Foundation, the U.S. Department of Energy, the National
Aeronautics and Space Administration, the Japanese Monbukagakusho, the
Max Planck Society, and the Higher Education Funding Council for
England. The SDSS Web Site is http://www.sdss.org/. 

The SDSS is managed by the Astrophysical Research Consortium for the
Participating Institutions. The Participating Institutions are the
American Museum of Natural History, Astrophysical Institute Potsdam,
University of Basel, University of Cambridge, Case Western Reserve
University, University of Chicago, Drexel University, Fermilab, the
Institute for Advanced Study, the Japan Participation Group, Johns
Hopkins University, the Joint Institute for Nuclear Astrophysics, the
Kavli Institute for Particle Astrophysics and Cosmology, the Korean
Scientist Group, the Chinese Academy of Sciences (LAMOST), Los Alamos
National Laboratory, the Max-Planck-Institute for Astronomy (MPIA),
the Max-Planck-Institute for Astrophysics (MPA), New Mexico State
University, Ohio State University, University of Pittsburgh,
University of Portsmouth, Princeton University, the United States
Naval Observatory, and the University of Washington. 

SMC acknowledges the support of an Australian Research Council QEII
Fellowship and an J G Russell Award from the Australian Academy of
Science.  GTR acknowledges support from an Alfred P. Sloan Research
Fellowship.  NPR and DPS acknowledge the support of National Science
Foundation grant AST06-07634.  MAS acknowledges the support of
National Science Foundation grant AST-0707266.

\vspace{1.0truecm}
This paper has been produced using the Blackwell Scientific Publications 
\TeX macros.

\end{document}